\documentclass[12pt]{article}
\usepackage{epsfig}
\newcommand{\be}{\begin{equation}}
\newcommand{\ee}{\end{equation}}
\newcommand{\ba}{\begin{eqnarray}}
\newcommand{\ea}{\end{eqnarray}}
\newcommand{\se}{\setcounter{equation}{0}}
\newcommand{\re}[1]{(\ref{#1})}

\newcommand{\1}{^{-1}}
\newcommand{\A}{{\cal A}}
\newcommand{\delt}{\delta^{\hspace*{-0.2mm}\mbox{\tiny G}}}
\newcommand{\dg}{^{\dagger}}
\newcommand{\di}{\mbox{d}}
\newcommand{\e}{\mbox{e}}
\newcommand{\E}{{\cal E}}
\newcommand{\f}{{\mbox{\scriptsize f}}} 
\newcommand{\g}{{\mbox{\scriptsize g}}} 
\newcommand{\bg}{\bar{g}} 
\newcommand{\bG}{\bar{G}} 
\newcommand{\G}{{\cal B}} 
\newcommand{\ga}{\gamma_5}
\newcommand{\h}{\frac{1}{2}}
\newcommand{\Id}{\mbox{1\hspace{-1.05mm}l}}   

\newcommand{\la}{\lambda}

\newcommand{\mo}[1]{^{(\mbox{\scriptsize #1})}}
\newcommand{\M}{{\cal M}}
\newcommand{\N}{{\cal N}}
\newcommand{\Op}{{\cal O}} 
\newcommand{\Pa}{{\cal P}} 
\newcommand{\bP}{\bar{P}} 
\newcommand{\ra}{\rightarrow}
 
\newcommand{\T}{{\cal T}} 
\newcommand{\Tr}{\mbox{Tr}} 
\newcommand{\bu}{\bar{u}} 
\newcommand{\ue}{|_{\U={\mbox{\scriptsize 1\hspace{-.7mm}l}}}} 
\newcommand{\U}{{\cal U}} 
\newcommand{\vp}{\varphi} 
\newcommand{\bw}{\bar{w}}
\newcommand{\W}{{\cal W}}

\begin{document}
\renewcommand{\baselinestretch}{1.0} \small\normalsize

\hfill {\sc HU-EP}-03/36

\vspace*{1cm}

\begin{center}

{\Large \bf General chiral gauge theories on the lattice} 

\vspace*{0.9cm}

{\bf Werner Kerler}

\vspace*{0.3cm}

{\sl Institut f\"ur Physik, Humboldt-Universit\"at, D-12489 Berlin, 
Germany}

\end{center}

\vspace*{1cm}

\begin{abstract}
We still extend the large class of Dirac operators decribing massless
fermions on the lattice found recently, only requiring that such operators
decompose into Weyl operators. After deriving general relations and 
constructions of operators, we study the basis representations
of the chiral projections. We then investigate correlation functions of 
Weyl fermions for any value of the index, stressing the related conditions
for basis transformations and their consequences, and getting the precise 
behaviors under gauge transformations and CP transformations. Various further
developments include considerations of the explicit form of the effective
action and of a representation of the general correlation functions
in terms of alternating multilinear forms. For comparison we also 
consider gauge-field variations and their respective applications.
Finally we compare with continuum perturbation theory.
\end{abstract}

\vspace*{0.8cm}

\section{Introduction}\label{INT}

We reconsider chiral gauge theories on the lattice generalizing the basic 
structure which has been introduced in the overlap formalism of Narayanan 
and Neuberger \cite{na93} and in the formulation of L\"uscher \cite{lu98}. 
The main aim of the generalization is to reveal the really relevant features 
and thereby to allow further developments of the subject. 

Massless Dirac operators $D$ on the lattice are functions of a basic unitary 
operator $V$.  The simplest case of this are Ginsparg-Wilson (GW) fermions 
\cite{gi82} for which $D$ is $\Id-V$ times a constant.  More generally this 
holds for the large class of operators 
\cite{ke02} where $D=F(V)$ satisfies $D+D\dg V=0$. In addition to GW fermions 
this class includes the ones proposed by Fujikawa \cite{fu00} and the 
extension of the latter \cite{fu02} as special cases. Here we go still 
further, only requiring that $D$ allows a decomposition into Weyl operators.

The chiral projections for this decomposition, which are implicit in 
Ref.~\cite{na93} and formulated in Ref.~\cite{lu98} in the GW case, are 
$P_-=\h(1-\ga V)$ and $\bP_+=\h(1+\ga)$. These forms have turned out to be 
suitable for the general class of operators in Ref.~\cite{ke02}, too.  In 
the GW case Hasenfratz \cite{ha02} has pointed that, instead of $\ga V$ and 
$\ga$, one could use $\ga((1-s)\Id+sV)/\N$ and $(s\Id+(1-s)V)\ga/\N$, 
respectively, with a parameter $s$. We here more generally introduce 
$\ga G(V)$ and $\bG(V)\ga$, respectively, with appropriate functions 
$G(V)$ and $\bG(V)$, which leads to the more general requirement 
$D+D\dg\bG G=0$ for $D$. 

Starting from the spectral representation of $V$ we first determine basic
conditions on $D$ and relations for its index. We then get the details
of the Weyl-operator decomposition and find that $G$ and $\bG$ must be
generally different. Next from the spectral representations of the chiral
projections we obtain detailed information about their structure. 
Our general construction of Dirac operators \cite{ke02} as well as the
related realizations of $V$ are seen to extend to the larger class of 
operators here. 

After making sure about the transformation properties of our general 
operators and a study of basis representations of chiral projections, 
which includes basis transformations, (finite) gauge transformations 
and CP transformtions, we give a formulation of the correlation functions 
of Weyl fermions for any value of the index. The additional conditions, which 
follow from the requirement that these functions must remain invariant
under basis transformations, are carefully discussed. The crucial meaning of 
the emerging decomposition of the total set of bases into subsets is stressed.

Considering gauge transformations the importance of finite transformations
in the analysis becomes apparent.  The general fermionic correlation 
functions exhibit gauge-covariant behavior of the fermion fields. In the 
exceptional cases, where either $G$ or $\bG$ equals the identity, in 
addition constant phase factors occur. The behavior under CP 
transformations turns out to differ from that of continuum theory 
by an interchange of $G$ and $\bG$, where the interchanged choice is a 
legitimate one, too. The effects of this interchange are also discussed.

Turning to further aspects we derive the explicit form of the effective 
action, give a formulation not referring to bases and address 
locality properties. Using the spectral representations we get a form of the 
correlation functions for general index and zero modes with a reduced 
chiral determinant. We next observe that the general correlation functions 
can be reformulated so that they are completely determined by alternating 
multilinear forms and $D$. Particular features of this representation are 
pointed out.

Also considering variations of the gauge field we discuss their application 
to specify basis-independent quantities. The properties of the variations
related to gauge transformations are obtained from those of the finite 
transformations and are seen to rely entirely on the latter. Considerations 
of variations of the effective action allow various comparisons. 

The developments in Ref.~\cite{lu98} and the investigations of CP properties
in Ref.~\cite{fu02} are discussed in the light of our results. A comparison
with continuum perturbation theory includes the discussion of related
Ward-Takahashi identities and the derivation of perturbative results on
the basis of the present nonperturbative definitions.

In Section 2 we introduce basic conditions and general relations. In Section 
3 we describe the realizations of the operators. Transformation properties 
of operators are considered in Section 4 and basis representations of the 
chiral projections are studied in Section 5. In Section 6 we investigate the
properties of correlation functions. Section 7 describes the 
alternating-form representation. Gauge-field variations are considered in 
Section 8. Section 9 is devoted to discussions of literature. The comparison 
with continuum perturbation theory is performed in Section 10. Section 11 
contains our conclusions.

\section{General conditions and relations}\se

\subsection{Basic unitary operator}

We consider a finite lattice with periodic boundary conditions and 
dimensionless quantities. Throughout we make consequent use of the fact that 
the operators describing fermions can be considered as acting in a unitary 
space of finite dimension.
 
Imposing the conditions
\be
V\dg=V^{-1}=\ga V \ga
\label{vv}
\ee
on $V$ we obtain the spectral representation
\be
V=P_1^{(+)}+P_1^{(-)}-P_2^{(+)}-P_2^{(-)}+\sum_{k\;(0<\vp_k<\pi)}
\big(\e^{i\vp_k} P_k\mo{I}+\e^{-i\vp_k} P_k\mo{II}\big)
\label{specv}
\ee
in which the orthogonal projections 
satisfy 
\be
\ga P_j^{(\pm)}=P_j^{(\pm)}\ga= \pm P_j^{(\pm)}\,,\quad \ga P_k\mo{I}=
P_k\mo{II}\ga. 
\label{PPg}
\ee
The dimensions of the right-handed and left-handed eigenspaces are 
$N_{\pm}(1)=\Tr\,P_1^{(\pm)}$ for eigenvalue $1$ and 
$N_{\pm}(-1)=\Tr\,P_2^{(\pm)}$ for eigenvalue $-1$. From \re{PPg} one gets
$N_k=\mbox{Tr}\,P_k\mo{I}=\mbox{Tr}\,P_k\mo{II}$ for the dimensions of the
other eigenspaces and the relations
\ba
\mbox{Tr}\big(\ga P_1^{(\pm)}\big)=\pm N_{\pm}(1),\quad
\mbox{Tr}\big(\ga P_2^{(\pm)}\big)=\pm N_{\pm}(-1),\nonumber\\
\mbox{Tr}\big(\ga P_k\mo{I}\big)=\mbox{Tr}\big(\ga P_k\mo{II}\big)=0.
\hspace*{38mm}
\label{TP}
\ea
With this we obtain
\be 
\lim_{\zeta\rightarrow 0}\mbox{Tr}\Big(\ga\frac{-\zeta}{V-\Id-\zeta\Id} \Big)
=N_+(1)-N_-(1)\hspace*{7mm}
\label{INDa}
\ee
and also find
\be 
\lim_{\zeta\rightarrow 0}\mbox{Tr}\Big(\ga\frac{V-\Id}{V-\Id-\zeta\Id} \Big)
=N_+(-1)-N_-(-1). 
\label{INDb}
\ee
Addition up these relations gives the sum rule 
\be  
N_+(1)-N_-(1)+N_+(-1)-N_-(-1)=0.
\label{sum}
\ee

The spectral representation \re{specv} with \re{PPg} also gives
\be
\mbox{Tr}(\ga V)=N_{+}(1)-N_{-}(1)-N_{+}(-1)+N_{-}(-1).
\label{VNN}
\ee
Combining \re{VNN} and \re{sum} we have
\be
N_{+}(1)-N_{-}(1)=\h\mbox{Tr}(\ga V).
\label{INN0}
\ee

\subsection{Dirac operator}

With \re{specv} the  spectral representation of $D=F(V)$ becomes 
\ba
D=f(1)\big(P_1^{(+)}+P_1^{(-)}\big)+f(-1)\big(P_2^{(+)}+P_2^{(-)}\big)
\nonumber
\\ +\sum_{k\;(0<\vp_k<\pi)}\Big(f(\e^{i\vp_k})P_k\mo{I}+f(\e^{-i\vp_k})
P_k\mo{II}\Big),
\label{specd}
\ea
in which $D$ is characterized by the spectral function $f(\e^{i\vp})$. On
the latter we impose three conditions. Firstly we require
\be
f^*(v)=f(v^*),
\label{cond+}
\ee
by which $D$ gets $\ga$-Hermitian, $\ga D\ga=D\dg$. Conversely 
$\ga$-Hermiticity of $D$ implies \re{cond+}. According to \re{cond+} 
obviously $f(1)$ and $f(-1)$ are real. Secondly imposing
\be
f(1)=0,
\label{cond0}
\ee
since the eigenvalue $1$ of $V$ corresponds to the eigenvalue $0$ of $D$, 
one describes massless fermions. Thirdly we then must have
\be
f(-1)\ne0
\label{cond1}
\ee
to allow for a nonvanishing index of $D$. This is seen noting that the index
is given by
\be 
I=\lim_{\zeta\rightarrow 0}\mbox{Tr}\Big(\ga\frac{-\zeta}{D-\zeta\Id} \Big),
\label{IND}
\ee
which for $f(-1)=0$ would give $N_+(1)-N_-(1)+N_+(-1)-N_-(-1)$ and thus
according to \re{sum} would be always vanishing.

With the conditions on $f$ in place we obtain for the index of $D$ 
\be 
I=N_+(1)-N_-(1).
\label{INDN}
\ee
With this \re{sum} tells that a nonvanishing value of $I$ requires the the 
occurrence of the eigenvalue $-1$ of $V$ in addition to the eigenvalue $1$ 
of $V$. Thus \re{sum} is seen to corresponds to the sum rule found by Chiu 
\cite{ch98} in the GW case. Further with \re{INN0} and \re{INDN} we have 
\be
I=\h\mbox{Tr}(\ga V),
\label{INN}
\ee
which generalizes results of the overlap formalism \cite{na93} and of the GW 
case \cite{ha98,lu98i}.

\subsection{Weyl operator decomposition}

We define chiral projections by
\be
P_{\pm}=P_{\pm}\dg=\h\Big(\Id\pm\ga G\Big),\qquad\bP_{\pm}=\bP_{\pm}\dg
=\h\Big(\Id\pm \bG\ga\Big),
\label{PR} 
\ee
where  the operators $G(V)$ and $\bG(V)$ satisfy
\be
G\1=G\dg=\ga G\ga,\qquad\bG\1=\bG\dg=\ga\bG\ga,
\label{GG}
\ee
and require that the decomposition
\be
D=\bP_+DP_-+\bP_-DP_+
\label{DP}
\ee
into Weyl operators holds. This immediately leads to the relations 
\be
\bP_{\pm}DP_{\mp}= DP_{\mp}=\bP_{\pm}D.
\label{PD}
\ee 
To obtain the condition on $D$, $G$ and $\bG$ needed in order that \re{DP}
holds we insert \re{PR} into it and find
\be
\bG\ga D+D\ga G=0.
\label{CHg}
\ee
Because of the $\ga$-Hermiticity of $D$, $G$ and $\bG$ we can write \re{CHg}
as $\bG\dg D+D\dg G=0$, which involves only commuting operators. We thus
obtain the condition 
\be
D+D\dg\bG G=0,
\label{CH+}
\ee
which has to be satisfied by $D$. This is seen to 
generalize our corresponding condition $D+D\dg V=0$ in Ref.~\cite{ke02},
thus still enlarging the class of operators found there.

\subsection{Spectral representations of chiral projections}

With \re{specv} $G(V)$ gets the spectral representation
\ba
G=g(1)\big(P_1^{(+)}+P_1^{(-)}\big)+g(-1)\big(P_2^{(+)}+P_2^{(-)}\big)
\nonumber
\\ +\sum_{k\;(0<\vp_k<\pi)}\Big(g(\e^{i\vp_k})P_k\mo{I}+g(\e^{-i\vp_k})
P_k\mo{II}\Big),
\label{specg}
\ea
and $\bG(V)$ the analogous one with the function $g$ replaced by the function
$\bg$. According to \re{GG} the functions $g$ and $\bg$ satisfy
\be
|g|^2=1,\quad g^*(v)=g(v^*),\qquad |\bg|^2=1,\quad\bg^*(v)=\bg(v^*).
\label{gg+}
\ee
In terms of spectral functions condition \re{CH+} reads
\be
f+f^*\bg g=0.
\label{ch+}
\ee
For $v=1$ because of \re{cond0} this is satisfied for any $g$ and $\bg$.
However, for $v=-1$ according to \re{cond1} this leads to the 
requirement
\be
\bg(-1)=-g(-1),
\label{ch-}
\ee
which causes $\bG$ and $G$ to be generally different.

For the difference of the numbers $\bar{N}=\Tr\,\bP_+$ and $N=\Tr\,P_-$ of 
the Weyl degrees of freedom in $\bP_+DP_-$, using the spectral 
representations of $G$ and $\bG$ and condition \re{ch-}, we obtain 
$\bar{N}-N=\h(\bg(1)+g(1))I$. Thus in order to have
\be
\bar{N}-N=I,
\label{INNw}
\ee
we must put
\be
\bg(1)=g(1)=1.
\label{bg11}
\ee
It then similarly follows that
\ba
N=\h\Tr\,\Id-I,\quad\,\bar{N}=\h\Tr\,\Id\hspace*{11.5mm}\mbox{ for } g(-1)=
-\bg(-1)=+1,\nonumber\\
N=\h\Tr\,\Id,\hspace*{12mm}\bar{N}=\h\Tr\,\Id+I\quad\mbox{ for } g(-1)=
-\bg(-1)=-1.\,
\label{TPb}
\ea

The spectral representations of $\ga G$ and $\bG\ga$ now become
\ba
\ga G=P_1^{(+)}-P_1^{(-)}+g(-1)(P_2^{(+)}-P_2^{(-)})+\sum_k(P_k^{[+]}-
P_k^{[-]}),\nonumber\\\bG\ga=P_1^{(+)}-P_1^{(-)}-g(-1)(P_2^{(+)}-P_2^{(-)})+
\sum_k(\bP_k^{[+]}-\bP_k^{[-]}),\,
\label{Gspec}
\ea
where the orthogonal projections $P_k^{[\pm]}$ and $\bP_k^{[\pm]}$ are given by
\ba
P_k^{[\pm]}=\h\Big(1\pm g(\e^{i\vp_k})\ga\Big)P_k\mo{I}+
\h\Big(1\pm g(\e^{-i\vp_k})\ga\Big)P_k\mo{II},\nonumber\\
\bP_k^{[\pm]}=P_k\mo{I}\h\Big(1\pm\bg(\e^{i\vp_k})\ga\Big)+
P_k\mo{II}\h\Big(1\pm\bg(\e^{-i\vp_k})\ga\Big).\,
\label{Ppr}
\ea

The spectral representations of $P_-$ and $\bP_+$ then for 
$g(-1)=-\bg(-1)=\pm1$ are
\ba
P_-=P_1^{(-)}+P_2^{(\mp)}+\sum_kP_k^{[-]},\nonumber\\
\bP_+=P_1^{(+)}+P_2^{(\mp)}+\sum_k\bP_k^{[+]}.\,
\label{Pspec}
\ea
With \re{PPg} it is obvious that one has $\mbox{Tr}\,P_k^{[-]}=\mbox{Tr}\,
\bP_k^{[+]}=\mbox{Tr}\,P_k\mo{I}=\mbox{Tr}\,P_k\mo{II}=N_k$. From \re{Pspec}
it is seen that $N-N_-(1)=\bar{N}-N_+(1)$, reflecting the fact that solely 
the zero modes of $D$ produce $\bar{N}\ne N$. On the other hand one gets 
$\tilde{N}:\,=N-N_-(1)=\bar{N}-N_+(1)=\sum_kN_k+N_{\mp}(-1)$ for $g(-1)=
-\bg(-1)=\pm1$, exhibiting the impact of the latter choice on the dimension
$\tilde{N}$.

So far \re{CH+}, which in terms of spectral functions is expressed by 
\re{ch+}, and which is a consequence of $DP_-=\bP_+D$, has been considered as 
a condition on $D$. Conversely, for given $D$, it provides a relation between
$G$ and $\bG$ and thus between $P_-$ and $\bP_+$. Indeed, for the projections
\re{Ppr} in their spectral representations \re{Pspec} with \re{ch+} and 
\re{specd} one gets explicitly
\be
DP_k^{[-]}D\dg=|f(\e^{i\vp_k})|^2\bP_k^{[+]},\qquad 
D\dg\bP_k^{[+]}D=|f(\e^{i\vp_k})|^2P_k^{[-]}.
\label{DPD}
\ee
Of the other projections in \re{Pspec} $P_2^{(\mp)}$ is seen to be related 
to itself, while for $P_1^{(-)}$ and $P_1^{(+)}$ because of $f(1)=0$ by $D$ 
no relation is provided.

\section{Realizations of operators}\se

\subsection{Special cases from literature}

The form implicit in Ref.~\cite{na93} and given in Ref.~\cite{lu98} 
in our notation corresponds to choosing
\be
\bG=\Id,\qquad G=V
\label{1V}
\ee
with GW operators $V$. This has been extended in Ref.~\cite{ha02} to
\be
G=\Big((1-s)\Id+sV\Big)/\N,\quad\bG=\Big(s\Id+(1-s)V\Big)/\N,
\label{Gs}
\ee
\be
\N=\sqrt{\Id-2s(1-s)\big(\Id-\textstyle{\h}(V+V\dg)\big)},
\label{Nn}
\ee
with a real parameter $s$. With an eigenvalue $\e^{i\vp}$ of $V$ one gets 
$1-2s(1-s)(1-\cos\vp)\ge0$ for the respective eigenvalue of $\N^{\,2}$, which
becomes zero for $s=\h$ and $\vp=\pi$, so that this definition does not work 
for $s=\h$. 

The chiral projections used in Ref.~\cite{fu02} in our notation are given
by functions $G$ and $\bG$ of the special form \re{Gs} with the more general
operator 
\be
V=1-\rho\1 D\,\Psi\Big((2\rho)^{-2}D\dg D\Big),
\label{FU12}
\ee
where $\rho$ is a constant and the operator function $\Psi$ is subject to 
$\Psi(X)\dg=\Psi(X)$. The Dirac operators associated to \re{FU12} have been 
shown in Ref.~\cite{ke02} to be a special case of the general class there. 

For the particular form \re{Gs} we obtain
\be
\bG G= V.
\label{GGV}
\ee
Therefore according to \re{CH+} in this case 
\be
D+D\dg V=0
\ee
holds, which has been the basic condition on $D$ in Ref.~\cite{ke02}. Thus 
\re{Gs} is suitable for the whole class of operators there.

\subsection{Construction of Dirac operator}

Our construction of $D$ in Ref.~\cite{ke02} can be extended to the present
more general case. For this purpose we first note that \re{ch+} can be written
as
\be
\Big(i(\bg g)^{-\h}f\Big)^*=i(\bg g)^{-\h}f
\ee
so that the function $q=i(\bg g)^{-\h}f$ is real. Therefore $f$ has the form
\be
f(\e^{i\vp})=-ip(\vp)q(\vp),\quad q^*(\vp)=q(\vp),\quad p(\vp)=
\Big(\bar{g}(\e^{i\vp})g(\e^{i\vp})\Big)^{\h}.
\label{fqq}
\ee

Noting that with \re{gg+} we have $(p^2(\vp))^*=p^2(-\vp)$ we choose the sign 
such that
\be
p^*(\vp)=p(-\vp).
\label{*qq}
\ee
With this and \re{fqq} condition \re{cond+} gives
\be
q(-\vp)=-q(\vp),
\ee
so that $q(0)=0$ and $f(1)=0$ hold, as required by \re{cond0}. 

Further noting that $p^2(\vp+2\pi)=p^2(\vp)$ we choose the sign such that 
\be
p(\vp+2\pi)=-p(\vp),
\label{SY3}
\ee
which according to \re{fqq} implies
\be
q(\vp+2\pi)=-q(\vp).
\label{SY4}
\ee
This will allow us to have $q(\pi)\ne0$ and $f(-1)\ne0$ as required by 
\re{cond1}.

With these conditions the spectral function $f$ and thus $D$ can be 
constructed. They differ from the ones in Ref.~\cite{ke02} only in that 
instead of the general function $p(\vp)$, there its special case $\e^{i\vp/2}$
occurs. Therefore with respect to the function $q$ we can rely on the result 
there. Its basic form which satisfies the conditions is $\sin\!\frac{\vp}{2}$.
This can be multiplied by a real function $w(\cos\vp)$ provided that 
$w(-1)\ne0$ so that \re{cond1} remains respected. Further, given
a function $q$ which satisfies the conditions then also $h(q)$ does if
$h$ is a real odd function, which in addition is strictly monotonous so that
still \re{cond1} holds. The steps of multiplying by a function of $\cos\vp$
and of taking an odd function of the result could be repeated, which we do,
however, not consider here. We then have 
\be
q(\vp)=h\Big(\sin\!\frac{\vp}{2}\;w(\cos\vp)\Big),
\label{GE4}
\ee
where the real functions of real argument $w$ and $h$ satisfy
\be
w(-1)\ne0,
\label{W-1}
\ee
\be
h(-x)=-h(x),\quad h(x_2)>h(x_1)\;\;{\rm for}\;\;x_2>x_1.
\label{MON}
\ee
With \re{MON} also the inverse function $\eta(y)$ of $h(x)$ is defined and 
strictly monotonous, 
\be
\eta\big(h(x)\big)=x,\quad \eta(-y)=-\eta(y),\quad \eta(y_2)>\eta(y_1)\;\;
{\rm for}\;\;y_2>y_1, 
\label{ETA}
\ee
which we shall need in the realization of $V$.

With \re{GE4} and \re{fqq} we get the form
\be
f(\e^{i\vp})=-i\Big(\bar{g}(\e^{i\vp})g(\e^{i\vp})\Big)^{\h}
\,h\Big(\sin\!\frac{\vp}{2}\;w(\cos\vp)\Big),
\label{GAi}
\ee
which inserted into \re{specd} gives the Dirac operator $D=F(V)$,
\be
D=-i\Big(\bG(V)G(V)\Big)^{\h}\,H\bigg(\frac{1}{2i}\Big(V^{\h}-V^{-\h}\Big)\;
W\Big(\h(V+V\dg)\Big)\bigg)\,,
\label{GAO}
\ee
where the properties of the operator functions $H$ and $W$ correspond to
those of the functions $h$ and $w$, respectively, and where the signs of
the roots are to be taken as defined in the context of spectral functions.

Several types of concrete examples of \re{GAO} have been worked out in 
Ref.~\cite{ke02} and methods to obtain further nontrivial ones have been
presented there. Therefore we do not pursue this issue further here.

\subsection{Related form of basic unitary operator}

To specify $V$ explicitly we introduce the normalization-type definition
\be
V=-D_E\Big(\sqrt{D_E\dg D_E}\,\Big)\1,
\label{Vg}
\ee
\be
D_E=-iE\Big(\sum_{\mu}\gamma_{\mu}{\cal S}_{\mu}\Big)+E_{\rm I}
\bigg(E_{\rm II}\Big(\sum_{\mu}(\Id-{\cal C}_{\mu})\Big)-E_{\rm II}
\Big(\vartheta\Id\Big)\bigg),
\label{VgD}
\ee
\be
{\cal S}_{\mu}=\frac{1}{2i}(\U_{\mu}-\U_{\mu}\dg),\quad{\cal C}_{\mu}=
\frac{1}{2}(\U_{\mu}+\U_{\mu}\dg),\quad(\U_{\mu})_{n'n}=U_{\mu n}
\delta^4_{n',n+\hat{\mu}},
\ee
with the gauge-field operator $\U_{\mu}$ and where the properties of the 
operator function $E$ correspond to those of the real function $\eta$ in
\re{ETA}. The functions $E$, $E_{\rm I}$ and $E_{\rm II}$ are Hermitian
operator functions of  Hermitian argument. They are required to be odd and
strictly increasing. This slightly generalizes the respective form in 
Ref.~\cite{ke02}, which arises here putting $E_{\rm I}=E$ and $E_{\rm II}(X)
=rX$ with $r>0$ and $\vartheta\Id=-m/r$. In addition specializing to $E(X)=X$
leads to the Neuberger operator \cite{ne98}. Instead of \re{Vg} using the 
representation \cite{ki98}
\be
V=-D_E\,\frac{1}{\pi}\int_{-\infty}^{\infty}\di s\frac{1}
{D_E\dg D_E+s^2}
\label{Vg1}
\ee
one can avoid the square root of noncommuting operators.

To confirm $\ga$-Hermiticity of $D_E$ one has to consider the individual terms
in \re{VgD}. For the function with $\gamma_{\mu}$ there one has to use the 
spectral representation of its argument for this purpose. Then with 
$\ga$-Hermiticity of $D_E$ one gets that of $V$, too. 

Having $\ga$-Hermiticity of $D_E$ one can also introduce the generalized
Hermitian Wilson-Dirac operator ${\cal H}=\ga D_E$, for which one gets
\be
\ga V=-\epsilon({\cal H}),
\label{EPS} 
\ee
providing a further form of the definition of $V$. Since the Hermitian and 
unitary operator $\ga V$ can have only the eigenvalues $\pm1$, according to
\re{EPS} only positive and negative eigenvalues of ${\cal H}$ must occur. To 
exclude zero modes of ${\cal H}$ (and thus also of ${\cal H}^2=D_E\dg D_E$), 
bounds on the gauge field as in Ref.~\cite{lu98} may be introduced. With 
\re{INN} and \re{EPS} one gets 
\be
I=\h\Tr(\ga V)=-\h\Tr\,\epsilon({\cal H}),
\label{INNy}
\ee
showing that the index of $D$ is also given by the difference of the numbers 
of positive and negative eigenvalues of ${\cal H}$. This extends the view of 
the overlap formalism \cite{na93} to the more general operators here. 

Checking the continuum limit in the free case for the Fourier transform
$\tilde{W}(\cos\vp)$ of $W$ one gets the condition 
\be
\tilde{W}(-1)\ne0,
\label{tW-1}
\ee
with which because of the monotony of $E_{\rm I}$ and $E_{\rm II}$ doublers 
are suppressed for $0<\vartheta<2$. Condition \re{tW-1} corresponds
to the requirement $f(-1)\ne0$ in \re{cond1}, needed to allow for a 
nonvanishing index. Since we work with dimensionless lattice quantities, in 
the limit we have $\tilde{D}/a\ra \tilde{D}_{\rm cont}\,$. Because of 
$H(E(X))=X$, putting 
\be
\tilde{W}(1)=2|\eta(m)|,
\label{tW+1}
\ee 
the usual normalization of the continuum propagator is obtained.

\section{Transformations of operators}\se

\subsection{Gauge transformations}

Under gauge transformations  the gauge-field operator $\U_{\mu}$ transforms as
\be
\U_{\mu}'=\T\U_{\mu}\T\dg,\qquad\T=\e^{\G},\qquad\G\dg=-\G,\qquad[\ga,\G]=0, 
\label{TG}
\ee
\be
(\U_{\mu})_{n'n}=U_{\mu n}\delta^4_{n',n+\hat{\mu}},\quad\G_{n'n}=
B_n\delta^4_{n'n},\quad B_n=i\sum_{\ell}b_n^{\ell}T^{\ell},
\label{TNN}
\ee
where $T^{\ell}$ are Hermitian generators and $b_n^{\ell}$ is real (and
which gives $U_{\mu n}'=\e^{B_{n+\hat{\mu}}}U_{\mu n}\e^{-B_n}$). 

Considering the spectral representations of a normal operator $\Op$ and the 
related one of a function $\Phi(\Op)$ of it,
\be
\Op=\sum_k\la_kP_k,\qquad \Phi(\Op)=\sum_k\phi(\la_k)P_k,
\ee
from $\Op'=\T \Op\T \dg$ we get for the orthogonal projections $P_k$ that
$P_k'=\T P_k\T \dg$, which implies that also $\Phi(\Op)'=\T \Phi(\Op)\T \dg$.

According to this the transformations of the operator functions in $\re{VgD}$
can be traced back to those of their arguments using the individual spectral 
representations of these arguments. Then with the form $\re{Vg1}$ it is seen 
that $V$ transforms as 
\be
V'=\T V\T \dg.
\label{TRV}
\ee
Further, since the spectral representations \re{specd} and \re{specg} of $D$, 
$G$ and $\bG$ are based on the same projections as \re{specv}  of $V$, with 
\re{TRV} we also have
\be
D'=\T  D\T \dg,\qquad\quad G'=\T G\T\dg,\qquad\quad\bG'=\T \bG\T\dg.
\label{TRD}
\ee
With this, \re{PR} and $[\ga,\T]=0$ we then further obtain
\be
P_{\pm}'=\T P_{\pm}\T\dg,\quad\qquad\bP_{\pm}'=\T\bP_{\pm}\T\dg. 
\label{TRP}
\ee

\subsection{CP transformations}

The operators $V$, $D$, $G$, $\bar{G}$ transform under charge conjugation as
\be
\Op(\U^{\rm C})=C\1\Big(\Op(\U)\Big)^{\rm T}C,
\label{CC}
\ee
where T denotes transposition in full space, $C$ is the charge conjugation
matrix\footnote{Using Hermitian $\gamma$-matrices with $\gamma_{\mu}^{\rm T}=
(-1)^{\mu}\gamma_{\mu}$ for $\mu=1,\ldots,4$ we choose $C=\gamma_2\gamma_4$.
This implies $\ga^{\rm T}=\ga$ and $[\ga,C]=0$ for $\ga=\gamma_1\gamma_2
\gamma_3\gamma_4$.}
with $C\gamma_{\mu}C\1=-\gamma_{\mu}^{\rm T}$ and  $C^{\rm T}=-C$, and 
where $\U^{\rm C}=\U^*$. To see this we first note that \re{CC} is satisfied 
by the arguments of the operator functions in $\re{VgD}$. Thus requiring $C\1=
C\dg$ and using the individual spectral representations of these arguments it 
follows for $D_E$, too, and considering $\re{Vg1}$ also for $V$. Then because 
the spectral representations of $D$, $G$ and $\bar{G}$ are based on that of 
$V$ this holds also for these operators.

For the parity transformation of the operators $V$, $D$, $G$, $\bar{G}$
we similarly get
\be
\Op(\U^{\rm P})=\Pa\gamma_4\Op(\U)\gamma_4\Pa
\label{PP}
\ee
where $\Pa_{n'n}=\delta^4_{n'\tilde{n}}$ with $\tilde{n}=(-\vec{n},n_4)$
and where we define $U_{4n}^{\rm P}=U_{4\tilde{n}}$ and $U_{kn}^{\rm P}=U_{k,
\tilde{n}-\hat{k}}$ for $k=1,2,3$. 

Combining relations \re{CC} and \re{PP} we have for the CP transformations 
of the operators $V$, $D$, $G$, $\bar{G}$,
\be
\Op(\U^{\rm CP})=\W\Big(\Op(\U)\Big)^{\rm T}\W\dg,\qquad\quad
\W=\Pa\gamma_4C\dg,
\label{WW}
\ee
where $\W\dg=\W\1$. 

With \re{WW}, $\ga^{\rm T}=\ga$ and $[\ga,C]=0$ we obtain for the 
chiral projections $\bP_+$ and $P_-$
\be
\W\Big(\bP_+(\U)\Big)^{\rm T}\W\dg=P_-^{\rm CP}(\U^{\rm CP}),\qquad
\W\Big(P_-(\U)\Big)^{\rm T}\W\dg=\bP_+^{\rm CP}(\U^{\rm CP}),
\label{PRCP0} 
\ee
where the transformed projections are defined by
\be
P_-^{\rm CP}=\h\Big(\Id-\ga\bG\Big),\qquad\quad\bP_+^{\rm CP}
=\h\Big(\Id+G\ga\Big).
\label{PRCP} 
\ee
This obviously differs from the definitions of $P_+$ and $\bP_-$ in \re{PR} 
by an interchange of $G$ and $\bG$. Taking the trace in \re{PRCP0} it is 
seen that one gets $I^{\rm CP}=-I$ for the index. 

A crucial observation now is that, since in \re{CH+} only the product of $G$ 
and $\bG$ enters, the same Dirac operator is associated to the operators
in \re{PRCP} as to those in \re{PR}. Applying \re{WW} and \re{PRCP0} to the 
Weyl operator $\bP_+DP_-$ we therefore consistently get
\be
\W\Big(\bP_+(\U)\,D(\U)\,P_-(\U)\Big)^{\rm T}\W\dg=\bP_+^{\rm CP}(\U^{\rm CP})
\,D(\U^{\rm CP})\,P_-^{\rm CP}(\U^{\rm CP}).
\label{WCP} 
\ee
Clearly the interchanged choice of $G$ and $\bG$ produced by the transformation
is a legitimate one as well. It is, however, not possible to get the symmetric
situation known from the continuum, since here due to \re{ch-} $\bG$ and $G$
must be generally different.

In more detail from the spectral representations \re{Pspec} it is seen that 
the two possible choices $g(-1)=-\bg(-1)=\pm1$ are interchanged under the 
transformation. In view of the impact of these choices on the dimensions in 
\re{TPb}, one could think of always fixing one of those dimensions to 
$\h\Tr\,\Id$ to avoid a change. With respect to the other terms nothing 
appears to prevent one from putting $\bg(\e^{i\vp_k})=g(\e^{-i\vp_k})$ 
by which one gets $\bP_k^{[\pm]}=P_k^{[\pm]}$ and thus no change there.

\section{Basis representations of chiral projections}\se

\subsection{Introduction and transformation of bases}

Noting that in full fermion space the vectors are specified by indices $n$, 
$\beta$ and $\alpha$ being related to position space, Dirac space and 
gauge-group space, respectively, we abbreviate the combination 
$(n,\alpha,\beta)$ by the index $\sigma$. Basis vectors $u_j$ with 
$j=1,\ldots,N$, which describe the $N$ Weyl degrees of freedom, then can be
considered as rectangular matrices of form $u_{\sigma j}$ and rank $N$. 

\hspace{0mm}From a more general point of view $u$ provides a mapping from the 
space $\E_{\rm w}$ of the Weyl degrees of freedom to the subspace $\E_P$ of 
full fermion space on which $P_-$ projects,\footnote{Between $\E_P$ and 
$\E_{\rm w}$ the mapping is even unitary, see  e.g.~Ref.~\cite{ka66} for 
the definition of unitary operators acting between different spaces.}
which both have dimension $N$. The respective transformations back are 
provided by $u\dg$ (which outside of $\E_P$ maps to zero). Analogous 
considerations apply to $\bP_+$ and a related basis $\bu$.

With the indicated understanding basis representations of the chiral 
projections are introduced by the conditions
\be
P_-=uu\dg,\quad u\dg u=\Id_{\rm w},\qquad\quad\bP_+=\bu\bu\dg,\quad 
\bu\dg\bu=\Id_{\rm \bw},
\label{uu}
\ee 
where $\Id_{\rm w}$ and $\Id_{\rm \bw}$ are the identity operators in the 
spaces of the degrees of freedom of Weyl fermions $\E_{\rm w}$ and of 
Weyl anti-fermions $\E_{\rm \bw}$, repectively. 

While the choice of the bases is not unique, different ones of them must 
represent the same projection. Thus they are related by unitary 
transformations, 
\be
u^{(S)}=uS,\quad S\1=S\dg,\qquad\bu^{(\bar{S})}=\bu\bar{S},\quad\bar{S}\1=
\bar{S}\dg, 
\label{BTR0}
\ee 
so that one generally gets $P_-=u^{(S)}u^{(S)\dag}$ and $\bP_+=\bu^{(\bar{S})}
\bu^{(\bar{S})\dag}$.  Obviously $S$ and $\bar{S}$ operate within $\E_{\rm w}$ 
and $\E_{\rm \bw}$, respectively.

\subsection{Gauge transformations}

According to $P_-'=\T P_-\T\dg$ from \re{TRP}, given a solution $u$ which 
satisfies conditions $P_-=uu\dg$ and $u\dg u=\Id_{\rm w}$ in \re{uu}, then 
$\T u$ is a solution of the transformed conditions, $P_-'=u'u'\,\!\dg$ and 
$u'\,\!\dg u'=\Id_{\rm w}$. Furthermore, then also $\T uS$ with any $S$ from
\re{BTR0} is a solution of the latter, and inserting all possible $S$ one 
gets all such solutions. Analogous conclusions hold for $\bu$, so that we 
generally have the forms
\be
u'=\T uS,\quad\quad\bu'=\T\bu\bar{S}.
\label{GuS}
\ee

The cases $G=\Id$, $\bG\ne\Id$ and $G\ne\Id$, $\bG=\Id$ are exceptional in
that $\T$ commutes with $P_-$ and $\bP_+$, respectively, so that the 
gauge-field dependences of the respective bases are no longer restricted. 
To see for $\bG=\Id$ how this gets consistent with \re{GuS} one notes that
one can put
\be
\T\bu=\bu\bar{S}_{\T}\quad\mbox{ for }\quad[\T,\bP_+]=0, 
\label{TRA} 
\ee
which allows to trade $\T$ for the basis transformation with
\be
\bar{S}_{\T}=\bu\dg\T\bu.
\label{TRAS}
\ee
In this way $\T \bP_+\T\dg=\bP_+$ is realized within $\bP_+=\bu\bu\dg$.
The particular case $\bu'=\bu$ is seen to arise by choosing $\bar{S}=
\bar{S}_{\T}\dg$ in \re{GuS}. 

In the case $G\ne\Id$, $\bG\ne\Id$, where no trading \re{TRA} is possible, 
\re{GuS} can be considered as a combination of a basis transformation 
\re{BTR0} and the gauge transformation
\be
u'=\T u,\quad\quad\bu'=\T\bu\qquad\mbox{ for }\quad G\ne\Id,\quad\bG\ne\Id.
\label{Gu}
\ee
In the exceptional case $G\ne\Id$, $\bG=\Id$ for the bases $\bu$ one can 
start from a basis $\bu_{\rm c}$ which is independent of the gauge field and 
gets the other ones by basis transformations $\bu=\bu_{\rm c}\bar{S}$. 
Then instead of \re{Gu} one has
\be
u'=\T u,\quad\quad\bu_{\rm c}'=\bu_{\rm c}\qquad\mbox{ for }\quad G\ne\Id,
\quad\bG=\Id.
\label{GuE}
\ee
Combining basis transformations with the transformations \re{Gu} and 
\re{GuE} all possible bases are reached, with the important consequence
that actually the whole original set of bases is related to the whole 
transformed one. A simple equivalent view of this is that in the transformation
laws \re{Gu} and \re{GuE} each basis can be any one of the respective set.

\subsection{CP transformations}

Given a solution $u$ of the conditions \re{uu}, then according to \re{PRCP0}
$\W\bu^*$ is a solution of the CP transformed conditions. With analogous 
conclusions for $\bu$ we thus arrive (with the dependences $u(\U)$, $\bu(\U)$,
 $u^{\rm CP}(\U^{\rm CP})$, $\bu^{\rm CP}(\U^{\rm CP})$) at 
\be
u^{\rm CP}=\W\bu^*,\quad\qquad\bu^{\rm CP}=\W u^*,
\label{Gv}
\ee
solving the transformed conditions 
\be
\bP_+^{\rm CP}=\bu^{\rm CP}\bu^{{\rm CP}\dag},\quad \bu^{{\rm CP}\dag}
\bu^{\rm CP}=\Id_{\rm w},\quad\quad P_-^{\rm CP}=u^{\rm CP}u^{{\rm CP}\dag},
\quad u^{{\rm CP}\dag}u^{\rm CP}=\Id_{\rm\bw}
\ee
(in which the interchange of $G$ and $\bG$ in \re{PRCP} as compared to \re{PR}
is implicit). Combining the transformations \re{Gv} with basis transformations
 all possible bases are reached, so that actually the whole original set of 
bases is related to the whole transformed one. An equivalent view is that in 
the transformation law \re{Gv} each basis can be  any one of the respective 
set.

\section{Correlation Functions}\se

\subsection{Definitions and general relations}

Associating Grassmann variables $\chi_k$ and $\bar{\chi_j}$ to the $N$ degrees
of freedom of left-handed fermions and the $\bar{N}$ ones of right-handed 
anti-fermions, respectively, the fermion field variables $\psi_{\sigma}$ and 
$\bar{\psi}_{\sigma'}$ get
\be
\bar{\psi}=\bar{\chi}\bu\dg,\quad \psi=u\chi.
\label{PS0}
\ee
The fermion action then is given by
\be
S_{\f}=\bar{\psi}D\psi=\bar{\chi}M\chi,
\label{SF0}
\ee 
where the matrix $M$, which maps from $\E_{\rm w}$ to $\E_{\rm\bw}$, is 
\be
M=\bu\dg Du.
\label{MM0}
\ee 

For a given value of the index $I$ the numbers $N$ and $\bar{N}$ are already
both determined since according to \re{TPb} either $N$ or $\bar{N}$ gets the 
fixed value $\h\Tr\,\Id$ and with \re{INNw} one generally has $\bar{N}-N=I$. 
Fermionic correlation functions 
$\langle\chi_{i_1}\ldots\chi_{i_L}\bar{\chi}_{j_1}\ldots\bar{\chi}_{j_{
\bar{L}}}\rangle_{\f}$ then can be nonvanishing only for
\be
L-N=\bar{L}-\bar{N}=r,\qquad 0\le r\le\min(\bar{N},N),\qquad\bar{L}-L=
\bar{N}-N=I,
\label{BLL}
\ee
where $\min(\bar{N},N)=\h(g(-1)I-|I|)$. We define such nonvanishing 
functions by
\ba
\langle\chi_{i_{r+1}}\ldots\chi_{i_N}\bar{\chi}_{j_{r+1}}\ldots
\bar{\chi}_{j_{\bar{N}}}\rangle_{\f}\hspace*{75mm}\nonumber\\
=s_r\int\di\bar{\chi}_{\bar{N}} \ldots\di\bar{\chi}_1\di\chi_N
\ldots\di\chi_1\;\;\e^{-S_{\f}}\;\;\chi_{i_{r+1}}\ldots\chi_{i_N}
\bar{\chi}_{j_{r+1}} \ldots\bar{\chi}_{j_{\bar{N}}}\;\nonumber\\
=\frac{1}{r!}\sum_{j_1,\ldots,j_r=1}^{\bar{N}}\,\sum_{i_1,\ldots,
i_r=1}^N\epsilon_{j_1,\ldots,j_{\bar{N}}}\epsilon_{i_1,\ldots,i_N}
M_{j_1i_1}\ldots,M_{j_ri_r},\hspace*{22.5mm}
\label{COR0}
\ea
where $s_r$ is the sign factor $s_r=(-1)^{rN-r(r+1)/2}$.

With \re{COR0} using \re{PS0} we get for the fields $\psi$ and $\bar{\psi}$
\ba
\langle\psi_{\sigma_{r+1}}\ldots\psi_{\sigma_N}\bar{\psi}_{\bar{\sigma}_{r+1}}
\ldots\bar{\psi}_{\bar{\sigma}_{\bar{N}}}\rangle_{\f}\hspace*{70mm}\nonumber\\
=s_r\int\di\bar{\chi}_{\bar{N}} \ldots\di\bar{\chi}_1\di\chi_N
\ldots\di\chi_1\;\;\e^{-S_{\f}}\;\;\psi_{\sigma_{r+1}}\ldots\psi_{\sigma_N}
\bar{\psi}_{\bar{\sigma}_{r+1}} \ldots\bar{\psi}_{\bar{\sigma}_{\bar{N}}}\nonumber\\
=\frac{1}{r!}\sum_{j_1,\ldots,j_{\bar{N}}=1}^{\bar{N}}\,\sum_{i_1,\ldots,
i_N=1}^N\epsilon_{j_1,\ldots,j_{\bar{N}}}\epsilon_{i_1,\ldots,i_N}
M_{j_1i_1}\ldots M_{j_ri_r}\nonumber\hspace*{22.5mm}\\\bu\dg_{j_{r+1}
\bar{\sigma}_{r+1}} \ldots\bu\dg_{j_{\bar{N}}\bar{\sigma}_{\bar{N}}}
u_{\sigma_{r+1}i_{r+1}}\ldots u_{\sigma_Ni_N}.
\label{COR}
\ea

In Ref.~\cite{lu98} the question of different complex factors multiplying the
fermionic correlation functions for different values $I$ has been raised.
There is, however, no theoretical principle describing this. An explicit 
reason for these factors could be the $I$-dependence in \re{TPb}. In 
Ref.~\cite{ha02} the importance of such factors for the magnitude of fermion 
number violating processes has been stressed. In Refs.~\cite{su00,fu02} 
suggestions that the modulus of them could possibly be generally one have 
been made. The alternating-form representations to be introduced in Section 7
might even suggest that there are no such factors. 

With the question of the indicated factors somehow settled, more general 
fermionic correlation functions can readily be constructed as linear 
combinations of the functions we have introduced. The inclusion of the 
gauge fields and the definition of full correlation functions then is 
straightforward and needs not to be considered here.

\subsection{Basis transformations}

Requiring that the field variables $\psi$ and $\bar{\psi}$ are not affected 
by the basis transformations \re{BTR0} induces transformations of the 
Grassmann variables $\chi$ and $\bar{\chi}$, too,
\be
\chi^{(S\dg)}=S\dg\chi,\quad\bar{\chi}^{(\bar{S})}=\bar{\chi}\bar{S}.
\label{BTG}
\ee
This has the consequence that the fermionic integration measure transforms as
\be
\di\bar{\chi}_{\bar{N}}^{(\bar{S})}\ldots\di\bar{\chi}_1^{(\bar{S})}\,\di
\chi_N^{(S\dg)}\ldots\di\chi_1^{(S\dg)}
\;=\;{\det}_{\rm\bw}\bar{S}\;\;{\det}_{\rm w}S\dg\;
\di\bar{\chi}_{\bar{N}} \ldots\di\bar{\chi}_1\,\di\chi_N \ldots\di\chi_1,
\label{MS}
\ee
where ${\det}_{\rm\bw}$ and ${\det}_{\rm w}$ denote the determinants in
the spaces $\E_{\rm\bw}$ and $\E_{\rm w}$, respectively. Thus, in order
to get invariance of the correlation functions \re{COR} we have to impose 
the additional conditions
\be
{\det}_{\rm w}S=1,\qquad{\det}_{\rm\bw}\bar{S}=1,
\label{TpD}
\ee
i.e.~to restrict the basis transformations to unimodular ones.\footnote{
For general expectations different constant phase factors of their 
contributions would lead to different interference terms in the moduli of the 
amplitudes. In order that basis transformations in different contributions
cannot cause such an effect the determinants must be fixed to a universal 
constant, which without restricting generality can be choosen to be one.}

Conditions \re{TpD} have important consequences for the possible sets of 
bases. In the case of $u$ (analogous considerations apply to $\bu$) without
the restriction \re{TpD} the unitary transformations $S$ connect all bases of 
the unitary space $\E_P$ on which $P_-$ projects. The unimodular $S$ only 
connect subsets of the total set of these bases, so that the total set is
decomposed into subsets. Our formulation of the theory thus has to be 
restricted to one of such subsets. The task then is to choose the appropriate 
one of them. 

We note that starting from an arbitrary basis the particular spectral
decompositions of the chiral projections in \re{Pspec} can also be reached
by unimodular transformations. This is so because a unitary transformation
can be expressed as a product of a unimodular transformation and of the
identity operator times a phase factor. Thus within this respect no 
restrictions arise.

Our discusssions of transformation properties, so far given for the total set,
apply as well to the subsets of bases. Some more detailed considerations (see 
Subsection 6.3) are only needed in the exceptional cases for gauge
transformations.  Generally the rule is that the symmetries of the 
chiral projections give that of the bases, where the latter are only fixed 
up to unimodular basis transformations.
 
The remaining problem then is that formulations in different subsets are not
equivalent. The non-unimodular basis transformations, which transform between
inequivalent subsets, produce phase factors. Such a phase factor describes 
how the results of the formulation of the theory in one subset differ from 
those of the formulation of the theory in another subset which is 
inequivalent to the former one. Obviously a criterion is needed, telling 
which one of such subsets is describing physics.

\subsection{Gauge transformations}

So far the combinations of the gauge transformations \re{Gu} and \re{GuE}
with basis transformations \re{BTR0} have been recognized to constitute 
the gauge transformations of the whole set of bases to the whole transformed 
set. After imposing conditions \re{TpD} the combinations of the unimodular 
basis transformations with the gauge transformations \re{Gu} and \re{GuE} 
give the transformations of the whole subset to the  whole transformed
subset. In this context it is to be noted that in the exceptional cases the 
subset of bases related to the gauge-field independent chiral projection 
necessarily contains a gauge-field independent basis.

With the correlation functions being invariant under unimodular basis 
transformations, in the non-exceptional case it suffices to use \re{Gu} 
to get the transformation properties. Accordingly inserting \re{Gu} into 
\re{COR}, the correlation functions are seen to transform as 
\ba
\langle\psi_{\sigma_1'}'\ldots\psi_{\sigma_L'}'\bar{\psi}_{\bar{\sigma}_1'}'
\ldots\bar{\psi}_{\bar{\sigma}_{\bar{L}}'}'\rangle_{\f}'\hspace*{85mm}
\nonumber\\=\sum_{\sigma_1,\ldots,\sigma_L}
\sum_{\bar{\sigma}_1,\ldots,\bar{\sigma}_{\bar{L}}}
\T_{\sigma_1'\sigma_1}\ldots\T_{\sigma_L'\sigma_L}
\langle\psi_{\sigma_1}\ldots\psi_{\sigma_L}\bar{\psi}_{\bar{\sigma}_1}
\ldots\bar{\psi}_{\bar{\sigma}_{\bar{L}}}\rangle_{\f}\,
\T_{\bar{\sigma}_1\bar{\sigma}_1'}\dg\ldots
\T_{\bar{\sigma}_{\bar{L}}\bar{\sigma}_{\bar{L}}'}\dg\nonumber\\
\mbox{for }\quad G\ne\Id,\quad\bG\ne\Id.\hspace*{16mm}
\label{TpT}
\ea

In the exceptional case $G\ne\Id$, $\bG=\Id$ with the transformation \re{GuE}  
we can apply the trading \re{TRA} to $\bu_{\rm c}$ to get $\bu_{\rm c}=
\T\bu_{\rm c}\bar{S}_{\T}\dg$ (where, of course, \re{TpD} needs not to hold
for $\bar{S}_{\T}$). With this the form \re{TpT} is seen to be  
supplemented by the constant phase factor ${\det}_{\rm\bw}\bar{S}_{\T}=
{\det}_{\rm\bw}(\bu_{\rm c}\dg\T\bu_{\rm c})$. To calculate this factor we 
represent the determinant by \cite{ke84}
\be
{\det}_{\rm\bw}\bar{S}=(-1)^{\bar{N}}\sum_{r=1}^{\bar{N}}\frac{(-1)^r}{r!}
\sum_{\rho_1=1}^{\bar{N}-r+1}\ldots\sum_{\rho_r=1}^{\bar{N}-r+1}
\delta_{\bar{N},\,\rho_1+\ldots+\rho_r}\frac{\Tr_{\rm\bw}(\bar{S}^{\rho_1})}
{\rho_1}\ldots\frac{\Tr_{\rm\bw}(\bar{S}^{\rho_r})}{\rho_r} 
\label{DETT}
\ee
and note that with $[\T,\bP_+]=0$ and $\T=e^{\G}$ we have
\be
\Tr_{\rm\bw}(\bar{S}^{\rho})=\Tr_{\rm\bw}\Big((\bu_c\dg\T\bu_c)^{\rho}\Big)=
\Tr\Big((\bP_+\T)^{\rho}\Big)=\Tr\Big(\bP_+\e^{\rho\G\bP_+}\Big).
\label{GGc}
\ee
Since ${\cal E}_{\rm\bw}$ and ${\cal E}_{\bP}$ both have dimension $\bar{N}$, 
we can use \re{GGc} to replace $\Tr_{\rm\bw}(\bar{S}^{\rho})$ in \re{DETT} 
and with $\bP_+=\h(1+\ga)\Id$ get 
\be
{\det}_{\rm\bw}\bar{S}={\det}_{\bP}(\e^{\G\bP_+})=
\e^{{\rm Tr}(\G\bP_+)}=\e^{\h{\rm Tr}\,\G}.
\label{eB}
\ee
Analogously for $G=\Id$, $\bG\ne\Id$, \re{TpT} is to be multiplied by the the 
phase factor $\e^{-\h{\rm Tr}\,\G}$. Because with \re{TNN} we have in more 
detail $\h\Tr\,\G=2i\sum_{n,\ell}b_n^{\ell}\,\mbox{tr}_{\g}T^{\ell},$
where the trace tr$_{\g}$ applies to gauge-field space only, obviously the 
additional condition $\mbox{tr}_{\g}T^{\ell}=0$ is needed to get rid of
these extra factors.\footnote{Since these factors can be different
in different fermionic contributions, similarly as the constant ones discussed
in the context of basis transformation, they are required to be equal to one.
The condition for this, $\mbox{tr}_{\g}T^{\ell}=0$, holds in the Standard 
Model.}

\subsection{CP transformations}

The combination of the CP transformations \re{Gv} with basis transformations
\re{BTR0} that satisfy in addition \re{TpD} constitutes the CP 
transformation of the whole respective subset of bases to the whole 
transformed subset. Since the correlation functions are invariant under 
such basis transformations, it suffices to use \re{Gv} to derive their
CP-transformation properties.

Inserting \re{WW} and \re{Gv} we get for the matrix \re{MM0} 
\be
M^{\rm CP}(\U^{\rm CP})=M^{\rm T}(\U).
\label{MM1}
\ee
With \re{MM1} and \re{Gv} we obtain for the transformed form of \re{COR}
\ba
\frac{1}{r!}\sum_{j_1,\ldots,j_{\bar{N}}=1}^{\bar{N}}\sum_{i_1,\ldots,i_N=1}^N
\epsilon_{j_1,\ldots,j_{\bar{N}}}\epsilon_{i_1,\ldots,i_N}M_{j_1i_1}^{\rm CP}
\ldots M_{j_ri_r}^{\rm CP}\nonumber\hspace*{42mm}\\\bu^{{\rm CP}\dag}_
{j_{r+1}\bar{\sigma}_{r+1}} \ldots\bu^{{\rm CP}\dag}_{j_{\bar{N}}\bar{
\sigma}_{\bar{N}}} u_{\sigma_{r+1}i_{r+1}}^{\rm CP}\ldots u_{\sigma_Ni_N}^{
\rm CP}\nonumber\\=\frac{1}{r!}\sum_{j_1,\ldots,j_{\bar{N}}=1}^{\bar{N}}
\sum_{i_1,\ldots, i_N=1}^N\;\sum_{\bar{\sigma}_{r+1}',\ldots\bar{\sigma}_{
\bar{N}}'}\sum_{\sigma_{r+1}',\ldots\sigma_N'}\epsilon_{j_1,\ldots,j_{
\bar{N}}}\epsilon_{i_1,\ldots,i_N} M_{i_1j_1}\ldots M_{i_rj_r}
\nonumber\hspace*{12mm}\\ \W_{\bar{\sigma}_{r+1}'\bar{\sigma}_{r+1}}\dg
\ldots\W_{\bar{\sigma}_{\bar{N}}'\bar{\sigma}_{\bar{N}}}\dg u_{\bar{
\sigma}_{r+1}' j_{r+1}}\ldots u_{\bar{\sigma}_{\bar{N}}'j_{\bar{N}}}
\bu\dg_{i_{r+1}\sigma_{r+1}'}\ldots\bu\dg_{i_N\sigma_{N}'}\W_{\sigma_{r+1}
\sigma_{r+1}'}\ldots\W_{\sigma_{N}\sigma_{N}'}.
\label{CORc}
\ea
This shows that the correlation functions \re{COR} transform as
\ba
\langle\psi_{\sigma_1'}^{\rm CP}\ldots\psi_{\sigma_L'}^{\rm CP}\bar{\psi}_{
\bar{\sigma}_1'}^{\rm CP}\ldots\bar{\psi}_{\bar{\sigma}_{\bar{L}}'}^{\rm CP}
\rangle_{\f}^{\rm CP}\hspace*{80mm}\nonumber\\=\sum_{\sigma_1,\ldots,\sigma_L}
\sum_{\bar{\sigma}_1,\ldots,\bar{\sigma}_{\bar{L}}}\W_{\bar{\sigma}_1\bar{
\sigma}_1'}\dg\ldots\W_{\bar{\sigma}_{\bar{L}}\bar{\sigma}_{\bar{L}}'}\dg
\langle\psi_{\bar{\sigma}_1}\ldots\psi_{\bar{\sigma}_{\bar{L}}}
\bar{\psi}_{\sigma_1}\ldots\bar{\psi}_{\sigma_L}
\rangle_{\f}\,\W_{\sigma_1'\sigma_1}
\ldots \W_{\sigma_L'\sigma_L}.
\label{TpC}
\ea
It is to be remembered here that according to \re{PRCP} an interchange of 
$G$ and $\bG$ is inherent in this, which in Subsection 4.2 has been discussed
in detail.

\subsection{Case of index zero and chiral determinant}

In the special case $\bar{L}=L$, where one gets a nonvanishing function for 
$\bar{N}=N$ and $I=0$ only, \re{COR} can also be written in the form\footnote{ 
Note that $\epsilon_{j_1,\ldots,j_r}^{i_1,\ldots,i_r}=1,-1$ or $0$ if
$i_1,\ldots,i_r$ is an even, an odd or no permutation of $j_1,\ldots,j_r$, 
respectively, with the special case $\epsilon_{j_1,\ldots,j_N}\equiv
\epsilon_{j_1,\ldots,j_N}^{1,\ldots,N}$.}
\ba
\langle\psi_{\sigma_{r+1}}\ldots\psi_{\sigma_N}\bar{\psi}_{\bar{\sigma}_{r+1}}
\ldots\bar{\psi}_{\bar{\sigma}_N}\rangle_{\f}=
\int\prod_{l=1}^N(\di\bar{\chi_l}\di\chi_l)\;\e^{-S_{\f}}\;
\psi_{\sigma_{r+1}}\bar{\psi}_{\bar{\sigma}_{r+1}}\ldots
\psi_{\sigma_N}\bar{\psi}_{\bar{\sigma}_N}=\nonumber\\
\sum_{\sigma_{r+1}',\ldots,\sigma_N'}
\epsilon\,_{\sigma_{r+1}\ldots\sigma_N}^{\sigma_{r+1}'\ldots\sigma_N'}\;
(P_-D\1\bP_+)_{\sigma_{r+1}'\bar{\sigma}_{r+1}} 
\ldots(P_-D\1\bP_+)_{\sigma_N'\bar{\sigma}_N}\;\;{\det}_{\rm\bw w}M,
\label{CORe}
\ea
where the notation ${\det}_{\rm\bw w}$ indicates that the determinant here 
actually involves a matrix connecting the different spaces $\E_{\rm w}$ and 
$\E_{\rm\bw}$. Correspondingly for $M\1$ one has the slightly generalized 
defining relations $M\1 M=\Id_{\rm w}$ and $MM\1=\Id_{\rm\bw}$. 

While in the form \re{COR} the presence of zero modes of $D$ is no problem,
in \re{CORe} one needs to care about them. If $M\1$ exists, using 
\re{specd} it follows from $M\1 M=\Id_{\rm w}$ that $P_1^{(-)}=0$ and from 
$MM\1=\Id_{\rm\bw}$ that $P_1^{(+)}=0$, so that $P_1^{(+)}+P_1^{(-)}$
vanishes and $D\1$ exists, too. Conversely, if $D\1$ exists, putting 
$M\1=u\dg D\1\bu$ it is seen that $M\1$ exists, too. Thus, if there are
zero modes of $D$ it follows that $M$ is not invertible, which implies 
${\det}_{\rm\bw w}M=0$ and also that $P_-D\1\bP_+=uM\1\bu$ does not exist. 

The basis independence of \re{CORe} becomes obvious noting that with \re{TpD}
one gets ${\det}_{\rm\bw w}(\bar{S}\dg MS)=({\det}_{\rm\bw}\bar{S}\dg)
({\det}_{\rm\bw w}M){\det}_{\rm w}S={\det}_{\rm\bw w}M$. The chiral 
determinant from \re{TpT} is seen to be gauge invariant for $\bG\ne\Id$, 
$G\ne\Id$, while in the exceptional cases $G\ne\Id$, $\bG=\Id$ and $G=\Id$,
$\bG\ne\Id$ the extra factors $\e^{\h{\rm Tr} \,\G}$ and $\e^{-\h{\rm Tr}\,
\G}$, respectively, occur. Under CP transformations because of \re{MM1} one 
has ${\det}_{\rm w\bw}M^{\rm CP}(\U^{\rm CP})={\det}_{\rm\bw w}M(\U)$.

\subsection{Effective action}

To evaluate the chiral determinant we write it as \cite{ke84}
\be
{\det}_{\rm\bw w}M=(-1)^N\sum_{r=1}^N\frac{(-1)^r}{r!}\sum_{\rho_1=1}^{N-r+1}
\ldots\sum_{\rho_r=1}^{N-r+1}\delta_{N,\,\rho_1+\ldots+\rho_r}
\frac{\Tr_{\rm\bw w}(M^{\rho_1})}{\rho_1}\ldots
\frac{\Tr_{\rm\bw w}(M^{\rho_r})}{\rho_r} 
\label{DETM}
\ee
and note that putting $Q=u\bu\dg$ we have
\be
\Tr_{\rm\bw w}(M^{\rho})=\Tr\Big((QD)^{\rho}\Big)=\Tr\Big(P_-(QD)^{\rho}P_-
\Big)=\Tr\Big(\bP_+(DQ)^{\rho}\bP_+\Big).
\label{TETM}
\ee
Since ${\cal E}_{\rm\bw}$, ${\cal E}_{\rm w}$, ${\cal E}_{\bP}$ and 
${\cal E}_P$ here all have dimension $N$, we can use \re{TETM} to replace 
$\Tr_{\rm\bw w}(M^{\rho})$ in \re{DETM} and obtain the factorization
\be
{\det}_{\rm\bw w}M={\det}_P(QD)={\det}_{\bP}(DQ)=
{\det}_{P\bP}(Q)\,{\det}_{\bP P}(D),
\label{DETF}
\ee
where ${\det}_{\bP P}(D)$ is the contribution of 
the Weyl operator $\bP_+DP_-$ while ${\det}_{P\bP}(Q)$ is 
the contribution of the bases. For the effective action we then have 
\be
\ln{\det}_{\rm\bw w}M=\Tr\ln Q+\Tr\ln(\bP_+DP_-).
\label{EFF}
\ee

The locality of the Weyl operator $\bP_+DP_-$ in \re{EFF} relies on that of
$D$, $P_-$ and $\bP_+$, which inherit locality from $V$. To study this the 
spectral representation of $V$ and the related ones of $F(V)$, $G(V)$ and 
$\bG(V)$ can be used. Considering the continuum analogon $V(x,y)=\sum_kv_k
\phi_k(x)\phi_k\dg(y)$ for this we have, for example, $D(x,y)=\sum_kf(v_k)
\phi_k(x)\phi_k\dg(y)$. Then considering the decrease of $\phi_k(y)$ and 
the orthogonality of the individual terms we see how the locality transfers.

For local $P_-=uu\dg$ and  $\bP_+=\bu\bu\dg$ the operator $Q=u\bu\dg$ in 
\re{EFF} gets local, too. To see this we consider the continuum analogues 
$P(x,y)=\sum_ku_k(x)u_k\dg(y)$, $\bP(x,y)=\sum_k\bu_k(x)\bu_k\dg(y)$ and 
$Q(x,y)=\sum_ku_k(x)\bu_k\dg(y)$. For appropriate decreasing of $P(x,y)$ 
and $\bP(x,y)$ for large $|y|$ also the individual terms for different $k$
decrease since they correspond to projections which are orthogonal to each
other. Because this means decreasing of $u_k(y)$ and $\bu_k(y)$, it
becomes obvious that also $Q(x,y)$ does appropriately decrease.
   
The operator $Q$ related to the basis contribution in the effective action
\re{EFF} obviously satisfies 
\be
QQ\dg=P_-,\qquad Q\dg Q=\bP_+,\qquad P_-Q=Q\bP_+=Q
\label{QQ}
\ee
and describes a unitary mapping between the spaces $\E_{\bP}$ and 
$\E_P$ for $\bar{N}=N$. We now observe that considering \re{QQ} as the
defining relations of $Q$, we can avoid referring to bases at all. 
The question of the inequivalent subsets of bases is replaced by that 
of a phase factor which is left open by \re{QQ}.

\subsection{General index and zero modes with determinant}

\hspace{0mm}From the spectral representations \re{Pspec} we have seen that
$\tilde{N}=N-N_-(1)=\bar{N}-N_+(1)$, which suggests
to introduce a $\tilde{N}\times\tilde{N}$ matrix $\tilde{M}$ from which
in contrast to $M$ the zero modes are removed. For this purpose we introduce
bases corresponding to the decomposition \re{Pspec}. Putting
$P_k^{[-]}=\sum_{j_k=1}^{N_k}u_{j_k}u_{j_k}\dg$ we see from \re{DPD} that 
\be
\bu_{j_k}=\e^{-i\Theta_k}|f(\e^{i\vp_k})|\1Du_{j_k}
\label{BU1}
\ee
with phases $\Theta_k$ gives the representation $\bP_k^{[+]}=\sum_{j_k=1}
^{N_k}\bu_{j_k}\bu_{j_k}\dg$ and because of
\be
D\dg D=|f(-1)|^2(P_2^{(+)}+P_2^{(-)})+\sum_{k\;(0<\vp_k<\pi)} 
|f(\e^{i\vp_k})|^2(P_k\mo{I}+P_k\mo{II})
\ee
and $P_k\mo{I}+P_k\mo{II}=P_k^{[+]}+P_k^{[-]}=\bP_k^{[+]}+\bP_k^{[-]}$
is properly normalized. Similarly we put $P_2^{(\mp)}=\sum_{j=1}^{N_{\mp}
(-1)}u_{j}u_{j}\dg$, which with
\be
\bu_{j}=\e^{-i\Theta}|f(-1)|\1f(-1)u_{j}=\e^{-i\Theta}|f(-1)|\1Du_{j}
\label{BU2}
\ee
and phase $\Theta$ leads to $P_2^{(\mp)}=\sum_{j=1}^{N_{\mp}(-1)}\bu_{j}
\bu_{j}\dg$.  The definition of the bases then is completed introducing 
$u_{j_-}$ and $\bu_{j_+}$ with $P_1^{(-)}=\sum_{j_-=1}^{N_{-}(1)}u_{j_-}
u_{j_-}\dg$ and $P_1^{(+)}=\sum_{j_+=1}^{N_{+}(1)}\bu_{j_+}\bu_{j_+}\dg$.

The nonvanishing matrix elements of \re{MM0} with the above bases and
identification of the indices are $M_{j_kj_k}=\e^{i\Theta_k}|f(\e^{i\vp_k})|$ 
and $M_{jj}=\e^{i\Theta}|f(-1)|$. We consider these elements as those of the 
diagonal $\tilde{N}\times\tilde{N}$ matrix $\tilde{M}$. Working out \re{COR0}
with this we obtain for it the form$^5$
\ba
\langle\chi_{i_{r+1}}\ldots\chi_{i_N}\bar{\chi}_{j_{r+1}}\ldots
\bar{\chi}_{j_{\bar{N}}}\rangle_{\f}=\frac{1}{(\tilde{N}-r)!}
\sum_{l_{r+1}',\ldots,l_{\tilde{N}}'}\sum_{l_{r+1},\ldots,l_{\tilde{N}}}
\tilde{M}_{l_{r+1}'l_{r+1}}\1\ldots\nonumber\hspace{10mm}\\\ldots
\tilde{M}_{l_{\tilde{N}}'l_{\tilde{N}}}\1
\;\epsilon_{l_{r+1}\ldots l_{\tilde{N}},\tilde{N}+1,\ldots,\bar{N}}
^{j_{r+1},\dots,j_{\bar{N}}}\;\;\epsilon_{l_{r+1}\ldots 
l_{\tilde{N}},\tilde{N}+1,\ldots,N}
^{k_{r+1},\dots,k_N}\;{\det}_{\tilde{N}}\tilde{M}
\label{COR1}
\ea
\be
{\det}_{\tilde{N}}\tilde{M}=\Big(\e^{i\Theta}|f(-1)|\Big)^{N_{\mp}(-1)}\prod_k
\Big(\e^{i\Theta_k}|f(\e^{i\vp_k})|\Big)^{N_k},
\ee
which with \re{PS0} gives
\ba
\langle\psi_{\sigma_{r+1}}\ldots\psi_{\sigma_N}\bar{\psi}_{\bar{\sigma}_{r+1}}
\ldots\bar{\psi}_{\bar{\sigma}_{\bar{N}}}\rangle_{\f}\hspace{79mm}\nonumber\\
=\sum_{\sigma_{r+1}',\ldots,\sigma_N'}\epsilon\,_{\sigma_{r+1}
\ldots\sigma_N}^{\sigma_{r+1}'\ldots\sigma_N'}\;\sum_{\bar{\sigma}_{r+1}',
\ldots,\bar{\sigma}_{\bar{N}}'}\epsilon\,_{\bar{\sigma}_{r+1}\ldots
\bar{\sigma}_{\bar{N}}}^{\bar{\sigma}_{r+1}'\ldots\bar{\sigma}_{\bar{N}}'}\;
\frac{1}{(\tilde{N}-r)!} (\tilde{P}_-\tilde{D}\1\tilde{\bP}_+)_{\sigma_{r+1}'
\bar{\sigma}_{r+1}'}\ldots\nonumber\\\ldots(\tilde{P}_-\tilde{D}\1
\tilde{\bP}_+)_{\sigma_{\tilde{N}}'\bar{\sigma}_{\tilde{N}}'}\;
u_{\sigma_{\tilde{N}+1},\tilde{N}+1}\ldots u_{\sigma_NN}\;
\bu_{{\tilde{N}}+1,\bar{\sigma}_{{\tilde{N}}+1}}\dg\ldots
\bu_{\bar{N}\bar{\sigma}_{\bar{N}}}\dg\;{\det}_{\tilde{N}}\tilde{M}
\label{CORx}
\ea
where $\tilde{P}_-$, $\tilde{D}$ and $\tilde{\bP}_+$ are the operators
$P_-$, $D$ and $\bP_+$ restricted to the subspace on which $\Id-P_1^{(+)}-
P_1^{(-)}$ projects.
While \re{CORx} resembles 
the conventional form \re{CORe}, in contrast to \re{CORe} it allows general
$N$ and $\bar{N}$ and zero modes.

Subsets of bases related to different choices of the phases $\Theta_k$ and 
$\Theta$ are obviously inequivalent if $\sum_kN_k\Theta_k+N_{\mp}(-1)\Theta$ 
gets different. The choices $\Theta_k=0$ and $\Theta=0$ in \re{CORx} 
corresponds to that in the generating functional of Ref.~\cite{fu02},
which has been derived using eigenfunctions of $DD\dg$. (This functional
does, however, not account for the restrictions related to the number 
of zero modes, which become explicit in our result \re{CORx}).

\section{Alternating-form representation}\se

\subsection{Structure of correlation functions}

Inserting \re{MM0} into \re{COR} we get the form
\be
\langle\psi_{\sigma_{r+1}}\ldots\psi_{\sigma_N}\bar{\psi}_{\bar{\sigma}_{r+1}}
\ldots\bar{\psi}_{\bar{\sigma}_{\bar{N}}}\rangle_{\f}
=\frac{1}{r!}\sum_{\bar{\sigma}_1\ldots\bar{\sigma}_r}\sum_{\sigma_1,\ldots,
\sigma_r}\bar{\Upsilon}_{\bar{\sigma}_1\ldots\bar{\sigma}_{\bar{N}}}^*
\Upsilon_{\sigma_1\ldots\sigma_N}D_{\bar{\sigma}_1\sigma_1}\ldots
D_{\bar{\sigma}_r\sigma_r},
\label{CORU}
\ee
with the totally antisymmetric quantities
\ba
\Upsilon_{\sigma_1\ldots\sigma_N}=\sum_{i_1,\ldots,i_N=1}^N\epsilon_{i_1, 
\ldots,i_N}u_{\sigma_{1}i_{1}}\ldots u_{\sigma_Ni_N},\nonumber\\
\bar{\Upsilon}_{\bar{\sigma}_1\ldots\bar{\sigma}_{\bar{N}}}=\sum_{j_1,\ldots,
j_{\bar{N}}=1}^{\bar{N}}\epsilon_{j_1,\ldots,j_{\bar{N}}}
\bu_{\bar{\sigma}_1j_1}\ldots\bu_{\bar{\sigma}_{\bar{N}}j_{\bar{N}}}.\,
\label{UPS}
\ea
For $\Upsilon_{\sigma_1\ldots\sigma_N}$ with $u$ and 
$\Upsilon_{\sigma_1\ldots\sigma_N}^{(S)}$ with $u^{(S)}=uS$ we obtain 
$\Upsilon_{\sigma_1 \ldots\sigma_N}^{(S)}=\Upsilon_{\sigma_1\ldots\sigma_N}
{\det}_{\rm w}S$, so that with \re{TpD} we have the basis independences
\be
\Upsilon_{\sigma_1 \ldots\sigma_N}^{(S)}=\Upsilon_{\sigma_1\ldots\sigma_N},
\qquad\bar{\Upsilon}_{\bar{\sigma}_1 \ldots\bar{\sigma}_{\bar{N}}}^{(
\bar{S})}= \bar{\Upsilon}_{\bar{\sigma}_1\ldots\bar{\sigma}_{\bar{N}}}.
\label{UPsu}
\ee

For $\bar{N}=N$, in an intermediate step using $\Tr_{\rm\bw w}((\bu\dg u)^{
\rho})=\Tr((Q)^{\rho})$ in the relation between determinant and traces, we 
obtain 
\be
\frac{1}{N!}\sum_{\sigma_1\ldots\sigma_N}\bar{\Upsilon}_{\sigma_1\ldots
\sigma_N}^*\Upsilon_{\sigma_1\ldots\sigma_N}={\det}_{P\bP}(Q)=\e^{{\rm Tr}
\ln Q},
\label{UPSx}
\ee
by which the contribution $\Tr\ln Q$ of the bases in in the effective action 
\re{EFF} is expressed solely in terms of $\Upsilon_{\sigma_1
\ldots\sigma_N}$ and $\bar{\Upsilon}_{\sigma_1\ldots\sigma_N}$.

It is instructive to compare the correlation function \re{CORU} of the 
chiral case with the ones of vector theory \cite{ke84}, 
\be
\langle\psi_{\sigma_{r+1}}\ldots\psi_{\sigma_K}\bar{\psi}_{\bar{\sigma}_{r+1}}
\ldots\bar{\psi}_{\bar{\sigma}_K}\rangle_{\f}^{\rm vect}
=\frac{1}{r!}\sum_{\bar{\sigma}_1\ldots\bar{\sigma}_r}\sum_{\sigma_1,\ldots,
\sigma_r}\epsilon_{\bar{\sigma}_1\ldots\bar{\sigma}_K}
\epsilon_{\sigma_1\ldots\sigma_K}D_{\bar{\sigma}_1\sigma_1}\ldots
D_{\bar{\sigma}_r\sigma_r}, 
\label{CORy}
\ee
in which $K=\Tr\,\Id$. It is seen that, while in the vector case one has the 
form $\epsilon_{\sigma_1\ldots \sigma_K}$ related to full fermion space, in
the chiral case one gets the forms $\Upsilon_{\sigma_1\ldots\sigma_N}$ and  
$\bar{\Upsilon}_{\bar{\sigma}_1\ldots\bar{\sigma}_{\bar{N}}}$ related to
its subspaces ${\cal E}_P$ and ${\cal E}_{\bP}$, respectively.

\subsection{Alternating multilinear forms}

An alternating multilinear form in $N$ variables is a scalar-valued function 
of $N$ vectors which is linear with respect to each of its arguments and 
vanishes if two of the arguments are equal. The latter implies the alternating,
i.e.~the change of sign if two of the arguments are interchanged. In the 
particular case, where $N$ is equal to the dimension of the respective vector 
space, such a form is completely determined \cite{di69} up to a scalar factor 
by its value for any set of bases taken as the arguments. 

This is exactly the situation of interest here, where $\Upsilon_{\sigma_1
\ldots\sigma_N}$ and $\bar{\Upsilon}_{\bar{\sigma}_1\ldots\bar{\sigma}_{
\bar{N}}}$ in \re{UPS} are seen to be explicit constructions of alternating 
multilinear forms in the spaces ${\cal E}_P$ and ${\cal E}_{\bP}$, 
respectively, the arguments of which are bases in $\sigma$-representation. 
The basis independence of these forms obviously realizes the general law. 

This suggests to introduce $\Upsilon_{\sigma_1\ldots\sigma_N}$ and 
$\bar{\Upsilon}_{\bar{\sigma}_1\ldots\bar{\sigma}_N}$, which together with
$D$ completely determine general correlation functions, in a slightly more
general way by  the relations
\be
\Upsilon_{\sigma_1\ldots\sigma_{i}\ldots\sigma_{j}\ldots\sigma_N}=
-\Upsilon_{\sigma_1\ldots\sigma_{j}\ldots\sigma_{i}\ldots\sigma_N},
\label{UPS1}
\ee
\be
\sum_{\sigma_i}(P_-)_{\sigma_j\sigma_i}
\Upsilon_{\sigma_1\ldots\sigma_i\ldots\sigma_N}=
\Upsilon_{\sigma_1\ldots\sigma_j\ldots\sigma_N},
\label{UPS2}
\ee
\be
\frac{1}{N!}\sum_{\sigma_1,\ldots,\sigma_N}\Upsilon_{\sigma_1\ldots\sigma_N}^*
\Upsilon_{\sigma_1\ldots\sigma_N}=1,
\label{UPS3}
\ee
for $\Upsilon_{\sigma_1\ldots\sigma_N}$ and by analogous ones for 
$\bar{\Upsilon}_{\bar{\sigma}_1\ldots\bar{\sigma}_{\bar{N}}}$.
The total antisymmetry of $\Upsilon_{\sigma_1\ldots\sigma_N}$ is imposed
by \re{UPS1}. The eigenequations \re{UPS2} determine it up to a 
normalization factor. The normalization then is fixed by \re{UPS3} 
up to a phase factor. The choice of the latter corresponds to the 
selection of one of the inequivalent subsets of bases considered before.

\section{Gauge-field variations}\se

\subsection{Definitions and general relations}

We define general gauge variations for a function $\phi(\U)$ by
\be
\delta\phi(\U)=\frac{\di\phi\big(\U(t)\big)}{\di t}\bigg|_{t=0}, \qquad 
\U_{\mu}(t)=\e^{t\G_{\mu}^{\rm left}}\U_{\mu}\e^{-t\G_{\mu}^{\rm right}},
\label{VAR}
\ee
where $t$ is a real parameter and where we have 
\be
(\G_{\mu}^{\rm left})_{n'n}=B_{\mu n}^{\rm left}\delta^4_{n',n},\quad\quad 
B_{\mu n}^{\rm left}=i\sum_{\ell}b_{\mu n}^{{\rm left}\ell}\,T^{\ell},
\label{TNx}
\ee
and analogous relations for $\G_{\mu}^{\rm right}$, with Hermitian generators 
$T^{\ell}$ and with $b_n^{{\rm left}\ell}$ and $b_n^{{\rm right}\ell}$ being 
real.  For the variations $\delt$ related to gauge transformations we get 
\be
\G_{\mu}^{\rm left}=\G_{\mu}^{\rm right} =\G,
\ee
with $\G$ as already met in \re{TG}.

\subsection{Variations of bases}

Varying the logarithm of ${\det}_{\rm w}S=1$ from \re{TpD} one gets the 
condition 
\be
\Tr_{\rm w}(S\dg\delta S)=0.
\label{dSS}
\ee
In \re{dSS} the restriction to the subset of bases connected by unimodular 
transformations of \re{TpD} is lost and an extension to the larger subset 
with constant phase factors for the determinant occurs.

We can rewrite \re{dSS} in terms of bases by inserting $S=u\dg u^{(S)}$ 
from \re{BTR0}, which gives
\be
\Tr(\delta u^{(S)}u^{(S)\dag})=\Tr(\delta u\,u\dg),
\label{condVU}
\ee 
showing that the quantity $\Tr(\delta u\,u\dg)$ is basis independent 
within the indicated extended subset. Obviously \re{condVU}, following from 
\re{dSS}, is analogous to \re{UPsu} for $\Upsilon_{\sigma_1\ldots\sigma_N}$ 
of the alternating-form representation, which follows from the original 
condition \re{TpD}.

We separate the inessential dependences on $\G_{\mu}^{\rm left}$ and 
$\G_{\mu}^{\rm right}$ off, getting
\be
\Tr(\delta u\,u\dg)=2i\mbox{ Im}\sum_{\mu,n}\mbox{tr}_{\g}
(\rho_{\mu n}\,\delta U_{\mu n}),
\label{RR}
\ee 
with the variation of the field
\be
\delta U_{\mu n}=B_{\mu,n+\hat{\mu}}^{\rm left}U_{\mu n}-U_{\mu n}
B_{\mu n}^{\rm right}
\label{RR1}
\ee 
and the quantity
\be
\rho_{\mu n,\alpha'\alpha}=\sum_{j,\sigma}u_{j\sigma}\dg
\frac{\partial u_{\sigma j}\hspace*{7mm}}{\partial U_{\mu n,\alpha\alpha'}},
\label{RRO}
\ee
which is invariant within the extended subset of bases. Conversely then 
$\rho_{\mu n}$ characterizes such a subset. It is to be noted that according 
to \re{TPb} $N$ may depend on $I$, in which case for each $I$ a different 
quantity $\rho_{\mu n}$ occurs. 

For the basis term in the effective action \re{EFF}, inserting $Q=u\bu\dg$,
we get
\be
\delta\,\Tr\ln Q=\Tr(Q\dg\delta Q)=
\Tr(\delta u\,u\dg)-\Tr(\delta\bu\,\bu\dg),
\label{QQu}
\ee
showing the relation to $Q$, which by \re{QQ} can also be defined
without referring to bases. Furthermore, with \re{UPSx} this can also be 
expressed in terms of $\Upsilon_{ \sigma_1\ldots\sigma_N}$ and $\bar{
\Upsilon}_{\sigma_1\ldots\sigma_N}$ of the alternating-form representation.

We add that also for the more general variations \re{VAR} the identity, 
used in Ref.~\cite{lu98},
\be
\delta_1\Tr(\delta_2u\;u\dg)-\delta_2\Tr(\delta_1u\;u\dg)
+\Tr(\delta_{[2,1]}u\;u\dg)=\Tr\Big(P_-[\delta_1P_-,\delta_2P_-]\Big),
\label{IDc}
\ee
follows readily from the defining relations \re{uu} of the bases (with the 
generators being $\G_{\mu(1)}^{\rm left}$, 
$\G_{\mu(1)}^{\rm right}$ and 
$\G_{\mu(2)}^{\rm left}$, $\G_{\mu(2)}^{\rm right}$ and $[\G_{\mu(2)}^{\rm 
left},\G_{\mu(1)}^{\rm left}]$, $[\G_{\mu(2)}^{\rm right},\G_{\mu(1)}^{\rm 
right}]$, respectively).

\subsection{Variations from gauge transformations}

For the operators $V$, $D$, $G$, $\bG$, $P_{\pm}$, $\bP_{\pm}$ gauge 
transformations in \re{TRV} -- \re{TRP} have been seen to have the explicit 
form ${\cal O}'=\T{\cal O}\T\dg$ where $\T =\e^{\G}$ and $\G=
\G_{\mu}^{\rm left}=\G_{\mu}^{\rm right}$. Introducing a real parameter $t$ 
the respective behavior can also be expressed by  
\be
{\cal O}(\U(t))=\T(t)\,{\cal O}(\U(0))\,\T\dg(t),\qquad\T(t)=\exp(t\G).
\label{OTO}
\ee
The variations $\delt$ corresponding to gauge transformations for
such operators therefore can be obtained simply by inserting \re{OTO} into 
\re{VAR}, with the result 
\be
\delt{\cal O}=[\G,{\cal O}].
\label{DEL}
\ee

With \re{DEL} we have $\delt P_-=[\G,P_-]$ for $P_-=uu\dg$, so that we get
$\delt u\,u\dg+u\,\delt u\dg=\G uu\dg-uu\dg\G$. This and the analogous 
relation for $\bP_+$ give the conditions
\be
(\delt u-\G u)u\dg+\big((\delt u-\G u)u\dg\big)\dg=0,\qquad
(\delt\bu-\G\bu)\bu\dg+\big((\delt\bu-\G\bu)\bu\dg\big)\dg=0,
\label{DuP}
\ee
which are to be satisfied by $\delt u$ and $\delt\bu$, however, are not
sufficient to determine them.

To obtain $\delt u$ and $\delt\bu$ we have again to resort to our knowledge 
from finite transformations. In the case $G\ne\Id$, $\bG\ne\Id$, proceeding 
similarly as above for operators, \re{Gu} can be expressed in the form
\be
u(\U(t))=\T(t)\,u(\U(0)),\qquad\bu(\U(t))=\T(t)\,\bu(\U(0)),
\label{TO}
\ee
which inserted into \re{VAR} gives 
\be
\delt u=\G u,\quad \delt\bar{u}=\G\bar{u}\qquad\mbox{ for }\quad G\ne\Id,\quad
\bG\ne\Id.
\label{DELu}
\ee
In the exceptional case $G\ne\Id$, $\bG=\Id$ with \re{GuE} we get
\be
\delt u=\G u,\quad \delt\bar{u}_c=0\qquad\mbox{ for }\quad G\ne\Id,\quad\bG=
\Id.
\label{DELuE}
\ee
It can be seen that \re{DuP} is satisfied by \re{DELu} and \re{DELuE} as it 
must be.  

It should be emphasized here that the results for the variations related to 
gauge transformations obviously rely entirely on the results for the finite 
transformations, for the operators as well as for the bases.

\subsection{Variations of effective action}

We obtain the variation of the effective action by varying our result 
\re{EFF}, which gives
\be
\delta\ln{\det}_{\rm\bw w}M=\Tr\Big(P_-D\1\bP_+\delta D+\delta u\,u\dg-
\delta\bu\,\bu\dg\Big).
\label{GA0}
\ee
For the last two terms in this we obtain with \re{RR} and the analogous 
relation for $\bu$
\be
\Tr(\delta u\,u\dg-\delta\bu\,\bu\dg)=
2i\mbox{ Im}\sum_{\mu,n}\mbox{tr}_{\g}
\Big((\rho_{\mu n}-\bar{\rho}_{\mu n})\,\delta U_{\mu n}\Big),
\label{RRR}
\ee 
indicating that $\rho_{\mu n}-\bar{\rho}_{\mu n}$ provides an 
additional term in the classical equation of motion. 

Specializing \re{GA0} to variations related to gauge transformations, with 
\re{DEL} inserted for $\delt D$, one gets 
\be
\delt\ln{\det}_{\rm\bw w}M=\Tr\Big((\bP_+-P_-)\G+\delt u\;u\dg-
\delt\bu\;\bu\dg\Big).
\label{GA}
\ee
For $G\ne\Id$, $\bG\ne\Id$, using \re{DELu} and \re{uu}, we obtain
\be
\Tr(\delt u\;u\dg)=\Tr(P_-\G),\qquad\Tr(\delt\bu\;\bu\dg)=\Tr(\bP_+\G),
\label{COMP}
\ee
so that in this case \re{GA} vanishes, as is to be expected from the result 
$\ln{\det}_{\rm\bw w}M'=\ln{\det}_{\rm\bw w}M$ for finite transformations.
For $G\ne\Id$, $\bG=\Id$, where according to \re{DELuE}
$\Tr(\delt u\;u\dg)=\Tr(P_-\G)$ and $\Tr(\delt\bu_c\;\bu_c\dg)=0$, because
of $\bP_+=\h(1+\ga)\Id$ we get
\be
\delt\ln{\det}_{\rm\bw w}M=\h\Tr\,\G,
\label{REE}
\ee
which is in perfect agreement with the result $\ln{\det}_{\rm\bw w}M'=
\ln{\det}_{\rm\bw w}M+\h\Tr\,\G$ for finite transformations. 

The term $\Tr((\bP_+-P_-)\G)$ in \re{GA} is the the gauge-anomaly term.
To see this in detail we insert \re{PR}, getting 
\be
\Tr\Big((\bP_+-P_-)\G\Big)=\h\Tr\Big(\ga\G(\bG+G)\Big),
\label{ANO}
\ee
and specialize to $\bG=\Id$, $G=V$, which with \re{TNN} gives 
\be
\Tr\Big((\bP_+-P_-)\G\Big)=
\h\Tr(\ga \G V)=i\sum_{n,\ell}b_n^{\ell}\,\h\mbox{tr}(\ga T^{\ell}V_{nn}).
\label{GAv}
\ee 
The term $\h\mbox{tr}(\ga T^{\ell}V_{nn})$ differs from the one of the 
chiral anomaly only by the insertion of the factor $T^{\ell}$. Since the 
inclusion of this factor in the derivation of the continuum limit is 
straightforward and because for the overlap $V$ this limit is safely known 
(see Ref.~\cite{ke01} for a proof and a discussion of literature) one gets
\be
\h\mbox{ tr}(\ga T^{\ell}V_{nn})\frac{1}{a^4}\ra -\frac{1}{32\pi^2}
\sum_{\mu\nu\la\tau}\epsilon_{\mu\nu\la\tau}\mbox{ tr}_{\g}\big(T^{\ell}
F_{\mu\nu}(x)F_{\la\tau}(x)\big).
\label{COr}
\ee
This still holds for the subclass of operators $D$ with $H(X)=X$ and any 
$W$ in \re{GAO} because only $V$ enters relation \re{INN}. It is also expected
for $H(X)=X^{2k+1}\,$, where the case of the chiral anomaly has been checked 
in Ref.~\cite{fu00a}. For more general $V$ and for other choices of $G$ and 
$\bG$ this limit remains to be investigated. 

The r.h.s~of \re{COr} vanishes if the anomaly cancelation condition 
$\mbox{tr}_{\g}(T^a\{T^b,T^c\})=0$ holds, which is crucial in continuum 
perturbation theory. In contrast to the latter, here for $\bG\ne\Id$, 
$\bG\ne\Id$ with \re{COMP} the bases provide a term compensating the 
anomaly term at the finite stage. In the exceptional cases 
$\bG\ne\Id$, $\bG=\Id$ and $\bG=\Id$, $\bG\ne\Id$ the compensation is
up to the constant $\h\Tr\,\G$ and $-\h\Tr\,\G$, respectively. 
This constant vanishes for $\mbox{tr}_{\g}\, T^{\ell}=0$ (which is satisfied
in the Standard Model).

\section{Discussions of literature}\se

\subsection{Formulation of L\"uscher}

In Ref.~\cite{lu98} the behavior of the effective action is investigated in
the special case of GW fermions. The chiral projections there in our
presentation correspond to the particular choice $G=V$, $\bG=\Id$ and the 
bases $\bu$ are restricted to ones independent of the gauge field. The form 
of the gauge-field variations there in our notation reads
$\delta U_{\mu n}=\eta_{\mu n}U_{\mu n}$. From \re{RR1} it is seen 
that with our general definition \re{VAR} one gets
\be
\eta_{\mu n}=B_{\mu,n+\hat{\mu}}^{\rm left}-
U_{\mu n}B_{\mu n}^{\rm right}U_{\mu n}\dg.
\label{BL}
\ee
Referring to linearity a current is defined there by putting 
\be
\Tr(\delta u\,u\dg)=-i\sum_{\mu,n}\mbox{tr}_{\g}(\eta_{\mu n}j_{\mu n}).
\label{CU}
\ee
In our presentation this current is explicitly given by 
\be
j_{\mu n}=i(U_{\mu n}\rho_{\mu n}+\rho_{\mu n}\dg U_{\mu n}\dg),
\label{VGrl}
\ee
where $\rho_{\mu n}$ is the quantity \re{RRO}. To specialize \re{CU} to the 
case of gauge transformations in our formulation means simply to put 
$B_{\mu n}^{\rm left}=B_{\mu n}^{\rm right}=B_n$, which 
gives
\be
\Tr(\delt u\;u\dg)=i\sum_n\mbox{tr}_{\g}\Big(B_n\sum_{\mu}(U_{\mu n}\dg
j_{\mu n}U_{\mu n}-j_{\mu,n-\hat{\mu}})\Big).
\label{CUg}
\ee
With this and \re{TNN} one gets for \re{GA} in the present special case the 
form
\be
\delt\ln{\det}_{\rm\bw w}M=\h\Tr(\ga \G V)+\Tr(\delt u\;u\dg)=i\sum_{n,\ell}
b_{n}^{\ell}X_{n\ell}
\label{CUl}
\ee
\be
X_{n\ell}=\h\mbox{ tr}(\ga T^{\ell}V_{nn})+
i\,\mbox{tr}_{\g}\Big(T^{\ell}\sum_{\mu}(U_{\mu n}\dg
j_{\mu n}U_{\mu n}-j_{\mu,n-\hat{\mu}})\Big).
\label{CUa}
\ee
where tr denotess the trace in gauge-field and Dirac space only.

The strategy in Ref.~\cite{lu98} was to impose appropriate conditions on
the current defined by \re{CU}, and then on the one hand side to look that it
determines a set of bases and on the other that such a current exists. The
relation to the subset of bases has been established requiring \re{IDc} for
the current after introducing it by \re{CU} into the terms there.\footnote{
This is a particular way to relate it to $P_-$. For the quantity $Q$ here
\re{QQ} is such a relation and for $\Upsilon_{\sigma_1\ldots\sigma_N}$ 
we have \re{UPS2} and for $\bar{\Upsilon}_{\bar{\sigma}_1\ldots
\bar{\sigma}_{\bar{N}}}$ an analogon thereof for this.} The existence so far 
could not be shown for the general nonAbelian case. 

The key quantity $\Tr(\delta u\,u\dg)$ of Ref.~\cite{lu98} is seen in \re{QQu}
to be the variation of $\Tr\ln Q$, where $Q$ can be defined by \re{QQ} 
without referring to bases or may be expressed by \re{UPSx} in terms of the 
quantities of the alternating-form representation. Thus instead of working 
with variations (which causes an extension of the subset of bases and needs 
additional smoothness properties) it is preferable to work with $Q$, which
as a unitary mapping between well defined spaces has also no existence
problem. 

Correspondingly also to work with the effective action itself is preferable,
which we did in Subsection 6.6. It is to be remembered, however, that 
considering the effective action only, means to restrict to $I=0$ and 
absence of zero modes of $D$. Furthermore $\Tr\ln Q$, and thus also its 
variation $\Tr(\delta u\,u \dg)$, do not cover the general case. This is 
seen from \re{UPSx}, in which $\Upsilon_{\sigma_1\ldots\sigma_N}$ and 
$\bar{\Upsilon}_{\bar{\sigma}_1\ldots\bar{\sigma}_{\bar{N}}}$ occur only 
for $\bar{N}=N$, while also $\bar{N}\ne N$ is needed in general correlation 
functions \re{CORU}. 

While the developments in Ref.~\cite{lu98} clearly represent a big step 
forward, the fact that properties of the chiral projections cause related 
properties of the bases, has not been sufficiently observed there. Within 
this respect the exclusive use of variations there has been a disadvantage. 
This holds, in particular, for gauge-transformation properties, the 
appropriate analysis of which here has been seen to need the use of finite 
transformations.

The discussion of gauge invariance in Ref.~\cite{lu98} has been based on 
\re{CUa}, presenting arguments that its contribution should vanish in the 
limit if the anomaly cancelation condition holds. Our result $\delt\ln{
\det}_{\rm\bw w}M=\h\Tr\,\G$ for the special case addressed there means 
that $X_{n\ell}=\h\mbox{tr}\,T^{\ell}$ holds without a further
condition at the finite stage.  

The problem that one has to select one of the subsets of bases out of the
inequivalent ones, which is describing physics, has been 
explained in Subsection 6.2. It has been pointed out to be still present in 
different form in the alternating-form representation with $\Upsilon_{\sigma_1
\ldots \sigma_N}$ and $\bar{\Upsilon}_{\bar{ \sigma}_1\ldots\bar{\sigma}_{
\bar{N}}}$ and in the formulation with $Q$ not referring to bases. It remains,
of course, there if instead of $\Tr\ln Q$ its variation $\Tr(\delta u\,u\dg)$
is used. In Ref.~\cite{lu98} this problem has not been addressed. In 
Ref.~\cite{lu00} arguments have been given that in perturbation theory the 
non-uniqueness of $\Tr(\delta u\,u\dg)$ would be irrelevant, while with
respect to the general non-perturbative case nothing could be concluded.

\subsection{CP investigations of Fujikawa, Ishibashi and Suzuki}

In \cite{ha02} Hasenfratz has observed that the divergence of \re{Gs} for 
$s=\h$ is an obstacle for getting the usual behavior under CP transformations.
Fujikawa et al.~\cite{fu02} have found that with their chiral projections,
which in our notation are given by functions $G$ and $\bG$ of the special 
form \re{Gs} with the more general operators $V$ from \re{FU12}, one 
encounters a singularity if one tries to enforce a symmetric situation like 
in the continuum. 

Here we have seen that actually $G$ and $\bG$ must be different because of 
very general reasons. It has turned out that to allow for a nonvanishing 
index of $D$ we have to impose the basic condition \re{cond1}. This together
with the defining condition $\re{CH+}$ for $D$ has lead to the requirement 
\re{ch-}, which generally forbids equality of the functions $G$ and $\bG$.
 
In the interesting investigations of CP properties of correlation functions 
in Ref.~\cite{fu02} the deviation from the continuum behavior found, in the 
notation of \re{Gs} is a replacement of $s$ by $1-s$. Obviously this is just 
the interchange of $G$ and $\bG$ which we find in the general case here.
We have seen that the interchanged choice belongs to the same $D$ and that
it is a legitimate one, too. 

The investigations of correlation functions in Ref.~\cite{fu02} are based on 
a generating functional which has been constructed using eigenfunctions of
$D\dg D$. The content of it is similar to that of \re{CORx} with the special
the choices $\Theta_k=\Theta=0$ (it does, however, not account for the 
restrictions related to the number of zero modes, which are explicit 
in our result \re{CORx}).  From our formulation it has 
become obvious that the change in the factor $|f(-1)|^{N_{\mp}(-1)}$ under 
the CP transformation is related to the two possible choices in \re{TPb}, as 
we have discussed together with further features at the end of Subsection 4.2. 

The generating functional constructed using eigenfunctions of $D\dg D$, in
Ref.~\cite{fu02} has been subject to a non-unimodular basis transformation,
considering the transformed form as the appropriate object. The motivation 
for this seems to be that the authors noticed the gauge invariance of their 
construction and that they wanted to accomodate the developments of
Ref.~\cite{lu98}. However, actually this is a transformation from one 
subset of bases, in which \re{TpD} had to be respected, to another subset 
of bases, in which \re{TpD} must be respected, too. With each of such subsets 
one cannot escape gauge covariance of the correlation functions\footnote{Apart
from the modification in the exceptional cases.}as we have seen 
here.\footnote{We add that it is $D\1\delta D$ which occurs in the variations 
of the reformulated basis term and of the Weyl term of the effective action as 
well, so that the non-locality argument in Ref.~\cite{fu02} does not apply.}

\section{Comparison with continuum perturbation theory}\se

\subsection{General observations}

That in the lattice formulation gauge invariance is obtained at the finite 
stage without the anomaly cancelation condition, which is crucial for 
renormalizability of continuum perturbation theory, is a remarkable 
observation. That in the exceptional cases a constant factor is produced 
unless the condition $\mbox{tr}_{\g}\,T^{\ell}=0$ holds, while this 
condition is not needed in continuum perturbation theory,\footnote{Though 
it is indeed satisfied in the Standard Model.} 
is a further difference. 

To assess these observations properly one has to note that at this point 
one actually is comparing rather different things. The formulation on the 
finite lattice is well defined in a unitary space. In continuum perturbation 
theory everything relies on the perturbation expansion with its definition in
the continuum. Since continuum perturbation theory is to be
taken as it is, the lattice approach has to be adapted for an adequate
comparison. This means that one has to derive the related continuum 
perturbation theory from the general lattice formulation of chiral gauge
theory and to compare the result with usual continuum perturbation theory.

In usual continuum perturbation theory the anomaly cancelation condition 
enters because appropriate Ward-Takahashi identities are needed in order 
that renormalizability can be established. These identities are related in
a certain way to gauge invariance. Since it is quite useful to see how
differences emerge also within this respect we compare the respective 
relations on the lattice and in the continuum before turning to the 
perturbation expansion. In this context we also clarify the r\^ole of
bases in the continuum case.

\subsection{Ward-Takahashi identities}

It suffices to restrict the considerations here to fermionic correlation 
functions. The Ward-Takahashi identities of interest on the lattice for 
vector theory are of form 
\be
\int[\di\bar{\psi}\di\psi] \e^{-S_{\rm f}}(\Op\,\delt_{\rm f}
S_{\rm f}-\delt_{\rm f}\Op)=0,
\label{WA0}
\ee
which follows from $\int[\di\bar{\psi}\di\psi]\e^{-S_{\rm f}} \Op$ by varying
only the fermion fields by a gauge transformation (for which one has $\delt_{
\rm f}\psi=\G\psi$, $\delt_{\rm f}\bar{\psi}=-\bar{\psi}\G$). This is actually 
just one example out of a variety of identities which on the lattice result 
from transformations leaving the integration measure invariant 
\cite{ka78,ka81,ke81}. 
It should be noted that the quantity $\delt_{\rm f}S_{\rm f}$ in \re{WA0} 
corresponds to a current derivative.\footnote{For example, for the Wilson 
action, putting $b_{n'}^{\ell'}=\delta^4_{n'n}\delta_{\ell'\ell}$ in $\G$ we 
obtain $\delt_{\rm f}S_{\rm f}=\sum_{\mu}(J_{\mu n}^{{\rm I}\,\ell}-
J_{\mu,n-\hat{\mu}}^{{\rm II}\,\ell})$ with $J_{\mu n}^{{\rm I}\,\ell}=
\frac{i}{2}\Big(\bar{\psi}_n(\gamma_{\mu}-r)U\dg_{\mu n}T^{\ell}
\psi_{n+\hat{\mu}}+\bar{\psi}_{n+\hat{\mu}}(\gamma_{\mu}+r)T^{\ell}U_{\mu n}
\psi_n\Big)$ and $J_{\mu n}^{{\rm II}\,\ell}=\frac{i}{2}\Big(
\bar{\psi}_n(\gamma_{\mu}-r)T^{\ell}U\dg_{\mu n}\psi_{n+\hat{\mu}}+
\bar{\psi}_{n+\hat{\mu}}(\gamma_{\mu}+r)U_{\mu n}T^{\ell}\psi_n\Big)$, which 
corresponds to $\sum_{\mu}{\cal D}_{\mu}J_{\mu}(x)$ of continuum theory 
with the adjoint representation form ${\cal D}_{\mu}$ of the covariant 
derivative. Furthermore, e.g.~for the choice $\Op=\psi_{n'}\bar{\psi}_{n''}$ 
in \re{WA0} one gets the lattice analogon of the familiar relation between 
vertex function and propagator.}

Since the validity of \re{WA0} for $\Op=\Id$ is a prerequisite for its
validity in the general case, in the following we consider this special case.
We note that for $\Op=\Id$ the underlying transformation can alternatively 
also be interpreted as a variation of the gauge field because one has 
\be
\int[\di\bar{\psi}\di\psi] \e^{-S_{\rm f}}\;\delt_{\rm f}S_{\rm f}=
-\delt\det D
\label{FG}
\ee
for $S_{\rm f}=\bar{\psi}D\psi$. Thus the relations
\be
\int[\di\bar{\psi}\di\psi] \e^{-S_{\rm f}}\;\delt_{\rm f}S_{\rm f}=0,
\label{FGcu}
\ee
\be
\delt\det D=0,
\label{FGde}
\ee
describing current conservation in the quantized theory and gauge invariance 
of the determinant, respectively, represent equivalent views.

In the continuum analogon of \re{FGcu}, with $S_{\rm f}=\int\di x^4\bar{\psi}\,
/\hspace*{-2.8mm}D\psi$ and the adjoint representation form ${\cal D}_{\mu}$
of the covariant derivative, 
\be
\int[\di\bar{\psi}\di\psi] \e^{-S_{\rm f}}\,\textstyle{\sum}_{\mu}
{\cal D}_{\mu}J_{\mu}=0,
\label{FGc}
\ee
the formal functional integral is defined by its perturbation expansion.
Correspondingly this expansion is to be checked. In this way one confirms 
the vanishing of \re{FGc}, i.e.~the conservation of the current 
$J_{\mu}^{\ell}(x)=\bar{\psi}(x)\gamma_{\mu}T^{\ell}\psi(x)$ in the quantized
case.

Switching to the chiral theory of the continuum with $S_{\rm f}=\int\di x^4
\bar{\psi}\,\h(1+\ga)/\hspace*{-2.8mm}D\psi$ the current in \re{FGc} becomes 
$J_{\mu}^{\ell}(x)=\bar{\psi}(x)\h(1+\ga)\gamma_{\mu}T^{\ell}\psi(x)$. Then 
checking of the expansion, by which the formal functional integral is defined, 
reveals the gauge anomaly in the triangle diagrams, so that current 
conservation in the quantized theory requires to impose the anomaly 
cancelation condition. 

We stress here that in this context actually also bases have to be 
introduced because the functional integration must be only over the occurring 
degrees of freedom. The reasons that the usual ignorance of this does not
spoil the results are firstly that in all integration variables trading for 
basis transformations is possible and secondly that in the fermion loops of 
the expansions the bases drop out. It appears of some importance to consider 
these issues here in more detail. 

The fermion fields are again of form $\bar{\psi}=\bar{\chi}\bu\dg$, 
$\psi=u\chi$, in which the integration variables $\bar{\chi}$, $\!\chi$ are
related to the degrees of freedom. The bases to be used here satisfy 
\be
u_{\rm co}u_{\rm co}\dg=\h(1-\ga)\Id,\qquad \bu_{\rm co}\bu_{\rm co}\dg=
\h(1+\ga)\Id.
\label{P55}
\ee
With respect to the variations of the fermion fields we note that for the 
gauge transformations $\psi'=\T\psi$ and $\bar{\psi}'=\bar{\psi}\T\dg$ now 
trading for basis transformations can be done for both $\psi$ and 
$\bar{\psi}$ (which in the exceptional cases on the lattice was only possible
either for $\psi$ or for $\bar{\psi}$). With \re{TRA} we get $\bar{\psi}'
=\bar{\chi}\bu_{\rm co}\dg\T\dg=\bar{\chi}\bar{S}_{\T}\dg\bu_{\rm co}\dg$ 
and similarly $\psi'=\T u_{\rm co}\chi=u_{\rm co}S_{\T}\chi$. Introducing new 
integration variables $\bar{\chi}'=\bar{\chi}\bar{S}_{\T}\dg$ and $\chi'=
S_{\T}\chi$ produces phase factors as calculated in \re{eB}, which here are
$\e^{-\h{\rm Tr}\,\G}$ and $\e^{\h{\rm Tr}\,\G}$, respectively. Since these
factors compensate each other the integration measure remains invariant.

In the fermion loops of the expansion the propagators are of form 
$u_{\rm co}\dg\,/\hspace*{-2.8mm}D_0\1\bu_{\rm co}$ and the vertices of form
$\bu_{\rm co}\dg\,\h(1+\ga)(/\hspace*{-2.8mm}D-/\hspace*{-2.8mm}D_0)u_{\rm co
}$, so that only the combinations $u_{\rm co}u_{\rm co}\dg$ and $\bu_{\rm co}
\bu_{\rm co}\dg$ occur, which according to \re{P55} are the chiral projections
and can be absorbed in the vertices. Thus the bases indeed drop out.

Turning now to the general formulation of the chiral theory on the lattice 
it is seen that the situation gets drastically different. The gauge field 
dependence of the bases no longer allows the trading as in the continuum 
(or in the exceptional cases not for both $\psi$ and $\bar{\psi}$). Thus 
nothing analogous to \re{FGcu} can be obtained and one remains with the 
condition corresponding to \re{FGde}. Furthermore, the bases no longer 
drop out from the loops of the perturbation expansion.

\subsection{Perturbation theory}

For the derivation of perturbation theory from the nonperturbative lattice
formulation we consider the form \re{CORe} of the correlation functions, 
discussing first the factors $P_-D\1\bP_+$ and then turning to the chiral
determinant ${\det}_{\rm\bw w}M$. It suffices again to restrict the 
considerations to the fermionic functions. 

For the discussion of the factors $P_-D\1\bP_+$, introducing $U_{\mu n}=
\exp(iag\sum_{\ell} A_{\mu n}^{\ell}T^{\ell})$ and the abbreviation 
$\A_s=ag A_{\mu n}^{\ell}$ with $s$ standing for the combination 
$(\mu,n,\ell)$, one can use the Taylor expansion
\be
f(\A)=f(0)+\sum_{k=1}^{\infty}\frac{1}{k!}\sum_{s_1,\ldots,s_k}\A_{s_1}\ldots
\A_{s_k}\;\bigg(\frac{\partial^k f(\A)}{\partial \A_{s_1}\ldots\partial 
\A_{s_k}}\bigg|_{\A=0}\bigg) 
\ee
to describe the gauge-field dependence of $V$. In the special case of the
overlap Dirac operator $\Id-V$ and of the dependence \re{Gs} of $G$
and $\bG$ on $V$ the explicit functions which occur are those obtained and 
used up to second order in Refs.~\cite{ki98,is99}. 
Because of $P_-D\1\bP_+=P_-D\1=D\1\bP_+$, in the exceptional cases 
one can use the combination with the constant projection so that nothing 
remains to be done. Otherwise one has to consider, for example, $P_-D\1=
\h(\Id-\ga G)D\1$ in more detail, which means to inspect the
limit of $GD\1$. Since with the free Dirac operator $D_0=D-D_{\rm I}$ we
have $D\1=D_0\1(\Id-D_{\rm I}D\1)$, this reduces to the consideration of
$GD_0\1$. We next remember that for the Fourier transform of $V$ in the 
free case one gets $\tilde{V}\ra1$ at zero and $\tilde{V}\ra-1$ at the 
corners of the Brillouin zone \cite{ke02}. We now note that actually 
still only such contributions survive in the limit if the gauge field is 
present. In the combination $VD_0\1$ then the corner contributions become 
suppressed so that one remains with $VD_0\1\ra D_0\1$. With this it is 
obvious that for the form \re{Gs} of $G$ one obtains $GD_0\1\ra D_0\1$ 
and thus altogether the correct limit for $P_-D\1\bP_+$. 

It is immediately obvious that these arguments extend to further
subclasses of operators constructed in Ref.~\cite{ke02} and also to further 
forms of $G(V)$ and $\bG(V)$ with appropriate dependences on $V$. A general
analysis for the whole class of operators within this respect remains to be 
performed.

To obtain the appropriate expansion\footnote{For our purpose it is not 
sufficient to make an ansatz for the effective action of form $\sum_{k=2}^{
\infty}\frac{1}{k!} \sum_{s_1,\ldots,s_k}\A_{s_1}\ldots\A_{s_k}\;{\cal V}_{
s_1,\ldots,s_k}$ as in Ref.~\cite{lu00}, but explicit details are to be 
worked out.} of ${\det}_{\rm\bw w}M$, we start from the decomposition 
\be
M=M_0+M_{\rm I},\qquad M_0=M\ue,
\ee
and with $D_0=D\ue$, $\bu_0=\bu\ue$ and $u_0=u\ue$ also have
\be
M_0=\bu_0\dg D_0 u_0,\qquad M_0\1=u_0\dg D_0\1\bu_0,\qquad \bP_{+0}=\bu_0
\bu_0\dg,\qquad P_{-0}=u_0u_0\dg.
\ee
For the chiral determinant we then get
\be
{\det}_{\rm\bw w}M=\int\prod_{l=1}^N(\di\bar{\chi_l}\di\chi_l)\;
\e^{-\bar{\chi}M\chi}=\Big(1+\sum_{\ell=1}^{\infty}z_{\ell}\Big)\;
{\det}_{\rm\bw w}M_0,
\label{Dper} 
\ee
\be
z_{\ell}=\frac{1}{\ell!}\sum_{j_1,\ldots,j_{\ell}=1}^N\sum_{k_1,\ldots,k_{
\ell}=1}^N\epsilon_{j_1,\ldots,j_{\ell}}^{k_1,\ldots,k_{\ell}}\;
(M_0\1 M_{\rm I})_{j_1k_1}\ldots(M_0\1 M_{\rm I})_{j_{\ell}k_{\ell}}.
\ee
The latter quantitiy can be written in the form \cite{ke84}
\be
z_{\ell}=\sum_{r=1}^{\ell}\frac{(-1)^{\ell+r}}{r!}\;\sum_{\rho_1=1}^{\ell-r+1}
\ldots\sum_{\rho_r=1}^{\ell-r+1}\delta_{\ell,\,\rho_1+\ldots+\rho_r}
\;\frac{t_{\rho_1}}{\rho_1}\ldots\frac{t_{\rho_r}}{\rho_r} 
\ee
with fermion loops given by 
\be
t_{\rho}=\Tr_{\rm ww}\Big((M_0\1M_{\rm I})^{\rho}\Big)=\Tr\Big((D_0\1\M)^{\rho}
\Big),
\label{tper} 
\ee
being made up of free propagators $D_0\1$ and vertices 
\be
\M=\bu_0M_{\rm I}u_0\dg.
\label{Mper}
\ee
With $D_{\rm I}=D-D_0$, $u_{\rm I}=u-u_0$ and $\bu_{\rm I}=\bu-\bu_0$ we 
can write the latter in detail as 
\be
\M=\bP_{+0}D_{\rm I}P_{-0}+\bu_0\bu_{\rm I}\dg Du_{\rm I}u_0\dg+\bu_0
\bu_{\rm I}\dg DP_{-0}+\bP_{+0}Du_{\rm I}u_0\dg.
\label{Mpe}
\ee

With respect to the term $\bP_{+0}D_{\rm I}P_{-0}$ in \re{Mpe} we note that  
to $D_0\1\bP_{+0}$ and to $P_{-0}D_0\1$ within \re{tper} analogous 
considerations as in the above discussion of $P_-D\1\bP_+$ apply, so that
in the limit $\bP_{+0}$ and $P_{-0}$ there can be replaced by $\h(1+\ga)$
and $\h(1-\ga)$, respectively. The fate of the other terms in \re{Mpe} in the 
limit of \re{tper} is determined by the behaviors of $u_{\rm I}$ 
and $\bu_{\rm I}$. Since to $V$ in the limit only the contributions stemming 
from zero and from the corners of the Brillouin zone survive, the respective 
chiral projections become independent of the gauge field (as in the 
exceptional cases anyway one of them is). Accordingly describing them in 
the limit by constant bases one gets $u_{\rm I}\ra0$ and $\bu_{\rm I}\ra0$.
Then only the term $\bP_{+0}D_{\rm I}P_{-0}$ of \re{Mpe} contributes to the 
limit and one gets the correct vertex function at tree-graph order.

The latter observation does not yet imply that \re{tper} gives the loops
of usual continuum perturbation theory in the limit of vanishing lattice
spacing $a\ra0$ because one-loop divergencies proportional to $1/a$ can 
compensate factors of $a$ and thus possibly lead to deviations from this. 
An example of such a phenomenon has been given in Ref.~\cite{ka78}. 
Therefore there remains the task for future investigations to clarify
whether and under which conditions one can arrive at usual perturbation
theory as desirable. 

A positive result of such investigations would mean that for the general 
formulation the anomaly cancelation condition is needed to preserve gauge 
invariance in the limit, too. With respect to the related mechanism it is
to be noted that on the lattice with all terms of $\M$ being present one 
has gauge invariance of \re{tper} (in the exceptional cases in addition 
requiring $\Tr\,T^{\ell}=0$) due to the compensation of the respective 
change of the first term on the r.h.s.~in \re{Mpe} by the changes of the 
terms there which involve $u_{\rm I}$ and $\bu_{\rm I}$. However, in case 
of the vanishing of the latter in the limit this invariance gets lost 
for $a\ra0$. Furthermore, the exceptional cases then are no longer 
distinct from the nonexceptional one.

The mentioned constancy of the bases in the limit is actually only needed
up to unimodular transformations. It has some impact on the selection of
bases discussed in Subsection~6.2. To see this we consider a basis $u$ 
(analogous considerations apply to $\bu$) of the general form 
$u=u_{\rm c}S$ where $u_{\rm c}$ is independent of the gauge field and 
$S$ a general unitary basis transformation. The constant bases can be 
represented by $u_{\rm c}=u_{\rm cb}S_{\rm c}$ where $u_{\rm cb}$ is one 
of them and the $S_{\rm c}$ are general constant basis transformations. 
The unitary transformations $S$ and $S_{\rm c}$ can be written as 
$S=\hat{S}\e^{i\phi/N}$ and $S_{\rm c}=\hat{S}_{\rm c}
\e^{i\phi_{\rm c}/N}$ where $\hat{S}$ and $\hat{S}_{\rm c}$ are unimodular 
and one has $\e^{i\phi}=\det S$ and $\e^{i\phi_{\rm c}}=\det S_{\rm c}$. The 
unimodular $\hat{S}$ and $\hat{S}_{\rm c}$ are irrelevant in ${\det}_{
\rm\bw w}M$ so that by putting $\hat{S}=\Id$ and $\hat{S}_{\rm cb}=\Id$ we 
can choose a convenient member of the respective subset (equivalence class).
We then are left with $u=u_{\rm cb}\e^{i(\phi_{\rm c}+\phi)/N}$ in which 
different values of $\phi_{\rm c}+\phi$ characterize different (inequivalent) 
subsets of bases. We see now that to get the vanishing of $u_{\rm I}$ the 
characteristic quantity $\phi$ of the selected subset must be independent 
of the gauge field.

\section{Conclusions}\se

We have still extended the large class of Dirac operators decribing massless
fermions on the lattice we have found recently, only requiring that such 
operators decompose into Weyl operators. Using the spectral representations
of the operators we have obtained a basic condition on the Dirac operator,
a general sum rule, a general expression for the index, a basic condition 
which prevents symmetry between the chiral projections and the detailed
structure of the chiral projections.  Our general construction of operators, 
using the tool of spectral functions, has been seen to extend, too. This 
also holds for the related realizations of the basic unitary operator, for 
which in addition some more freedom has been observed.

After making sure about the transformation properties of our general operators
and performing a careful study of the basis representations of the chiral 
projections, we have turned to the correlation functions of Weyl fermions. 
For their investigation we have introduced a formulation of the fermionic 
functions which works for any value of the index. For the additional 
conditions due to the requirement of their invariance under basis 
transformations the consequences have been made precise. In this context we 
have stressed that, since formulations in different ones of the emerging 
subsets of bases are not equivalent, a criterion is needed telling 
which one of such subsets describes physics. 
 
Considering gauge transformation the crucial importance of using finite 
transformations in the analysis has become obvious. We have seen that the 
correlation functions exhibit gauge-covariant behavior of the fermion
fields. In the exceptional cases in addition constant phase factors have 
turned out to occur, the values of which have been determined. 
The behavior under CP transformations has been found to differ from that 
known from continuum theory by involving an interchange of the functions in 
the chiral projections, where the interchanged choice is a legitimate one,
too. In view of our result that such functions must be generally different, 
we have also studied the effects of the interchange. 

We have derived the explicit form of the effective action, in which
the contributions of the Weyl operator and of the bases are separated. 
This form has allowed us to introduce a formulation not referring to bases
and to discuss locality properties. Starting from the observation that for 
any value of the index the removal of zero modes makes the remaining chiral 
matrix quadratic, we have used the spectral representations to get a form of 
general correlation functions with a reduced chiral determinant. 
We have reformulated the correlation functions so that they are 
completely determined by alternating multilinear forms and $D$ and
discussed the features of this presentation. 

Variations of the gauge fields have been defined with left and right 
generators. Their application to specify basis-independent quantities has
been discussed and the extension of the related subsets beyond that of the 
unimodular case noted. The properties of variations related to gauge 
transformations have been obtained from those of the finite transformations 
and been seen to rely entirely on the latter. Considerations of the
variations of the effective action have allowed various comparisons.

The developments in Ref.~\cite{lu98} and the investigations of CP properties 
of Ref.~\cite{fu02} have been discussed in the light of our results.
A comparison with continuum perturbation theory has included the discussion 
of related Ward-Takahashi identities, the derivation of perturbative 
results on the basis of the present nonperturbative definitions and the
discussion of the relevant conditions.

\section*{Acknowledgement}

I wish to thank Michael M\"uller-Preussker and his group for their kind 
hospitality.

\end{document}